\documentclass[sigconf,nonacm]{acmart}

\usepackage{colortbl}
\usepackage{tikz}
\usepackage{pgfplots}
\usepgfplotslibrary{groupplots}
\pgfplotsset{compat=1.18}
\usepackage{enumitem}
\usepackage{amsmath,amssymb,amsfonts}
\usepackage{verbatim}
\usepackage{float}
\usepackage{dblfloatfix}
\usepackage{caption}
\usepackage{amsmath,amsthm,amsfonts,amssymb,amscd,bm,amsbsy}
\usepackage{balance}
\usepackage[export]{adjustbox}
\usepackage{subfigure}
\usepackage{enumerate}
\usepackage{fancyhdr}
\usepackage{pifont}
\usepackage{mathrsfs}
\usepackage{xcolor}
\usepackage{graphicx}
\usepackage{listings}
\usepackage{float}
\usepackage{textcomp}
\usepackage{xcolor}
\usepackage{ragged2e}
\usepackage[utf8]{inputenc}
\usepackage{graphicx}
\usepackage{lipsum}
\usepackage{indentfirst}
\usepackage{hyperref}
\usepackage{cleveref}
\usepackage{tabularx}
\usepackage{balance}
\usepackage{wrapfig}
\usepackage{multirow}
\usepackage{array}
\usepackage{capt-of}
\usepackage{booktabs}

\usepackage{courier}
\usepackage[ruled, vlined, linesnumbered]{algorithm2e}
\usepackage{algpseudocode}
\graphicspath{{fig/}}
\makeatletter
\def\blfootnote{\gdef\@thefnmark{}\@footnotetext}
\makeatother
\usetikzlibrary{arrows.meta,bending}
\usepackage{newunicodechar}
\newunicodechar{①}{\textcircled{1}}
\newunicodechar{②}{\textcircled{2}}
\newunicodechar{③}{\textcircled{3}}
\newunicodechar{④}{\textcircled{4}}
\newunicodechar{⑤}{\textcircled{5}}
\newunicodechar{⑥}{\textcircled{6}}

\definecolor{gjr}{HTML}{FF0000}

\newcommand{\ourname}{InSituANN}

\newcommand{\pkuAffiliation}{
\affiliation{
  \institution{Peking University}
  \city{Beijing}
  \country{China}
}
}

\setcopyright{none}
\renewcommand\footnotetextcopyrightpermission[1]{}

\makeatletter
\renewcommand{\@fnsymbol}[1]{%
  \ensuremath{%
    \ifcase#1
    \or \dagger
    \or \mathsection
    \or \mathparagraph
    \or \|
    \else\@ctrerr
    \fi}}
\makeatother

\begin{document}

\author{Yuemeng Xu}
\authornote{Yuemeng Xu and Zongxi Liu made primary contributions to this work.}
\pkuAffiliation

\author{Zongxi Liu}
\authornotemark[1]
\pkuAffiliation

\author{Junyu Long}
\pkuAffiliation

\author{Yiming Huang}
\pkuAffiliation

\author{Jiarui Guo}
\pkuAffiliation

\author{Yangyujia Wang}
\pkuAffiliation

\author{Jiachen Xu}
\pkuAffiliation

\author{Dongyuan Yu}
\pkuAffiliation

\author{Zongwei Lv}
\pkuAffiliation

\author{Tong Yang}
\authornote{Tong Yang is the corresponding author.}
\pkuAffiliation
\renewcommand{\shortauthors}{Xu et al.}
\title{\ourname{}: Revisiting IVF for PCIe-Efficient Billion-Scale Vector Search}

\vspace{-0.35em}
\begin{abstract}
Approximate nearest neighbor search (ANNS) over billion-scale
vector datasets has become a foundational operator for modern
retrieval systems, powering large-scale recommendation, semantic
search, and LLM/RAG workloads. Although GPUs offer massive parallelism and high-bandwidth memory for batched vector search, their limited VRAM capacity makes fully
GPU-resident billion-scale indexes difficult to deploy. In
CPU--GPU heterogeneous designs, keeping the base vectors in host
memory avoids this capacity limit, but naively offloading fine search
to the GPU introduces a new bottleneck: large volumes of base-vector
data must be streamed over PCIe.

We present InSituANN, an IVF-based ANNS engine that enables
billion-scale vector search on a single commodity GPU. InSituANN
keeps original base vectors in host memory, performs fine search
in situ, and uses the GPU for compact routing and optional pruning.
As a result, query processing avoids PCIe transfers of high-dimensional
base vectors while retaining the simplicity of IVF.
Beyond query performance, we further design an ultra-fast IVF construction path for \ourname{}. On SIFT-1B, \ourname{} builds the IVF index
in 5.2 minutes, about $350\times$ faster than the measured 30.4-hour HNSW
build. At matched recall on billion-scale datasets, \ourname{} improves
end-to-end throughput by 104.9$\times$--4298.2$\times$ over the PCIe-bound Rummy
baseline and by $2.4\times$--$4.6\times$ over DiskANN on SIFT-1B and DEEP-1B.
Together with strong recall--throughput trade-offs and lower index space than
graph-based alternatives, these gains make billion-scale retrieval practical
on cost-efficient hardware.

We open-source \ourname{} at \url{https://github.com/mindtravel/InSituANN-OpenSource}.
\end{abstract}

\maketitle

\vspace{-0.65em}

\setlength{\subfigcapskip}{-0.15cm}
\setlength{\subfigbottomskip}{-0.05cm}

\sloppy

\section{Introduction}
\label{sec:introduction}

Approximate nearest-neighbor (ANN) search avoids exhaustive scans by trading a
small loss in accuracy for substantially higher throughput over million- to
billion-scale vector collections~\cite{arya1993approximate,datar2004locality,malkov2018efficient}.
It lies on the critical path of retrieval-augmented generation, recommendation,
semantic search, and multimodal retrieval~\cite{lewis2020retrieval,jing2025large,covington2016youtube,li2017fexipro,karpukhin2020dpr,radford2021clip}.
As embedding collections grow, ANN efficiency increasingly determines
end-to-end latency, throughput, and system scalability~\cite{wang2021milvus,sun2025gaussdb,Wei2020AnalyticDBV,jing2025large}.

Among ANN designs, graph-based indexes such as HNSW and DiskANN are widely
adopted for high-recall search~\cite{malkov2018efficient,jayaram2019diskann,manohar2024parlayann}.
Their query-dependent pointer chasing over proximity graphs produces irregular
memory accesses that can degrade cache locality and increase latency
variability, especially on GPUs and tiered-memory systems~\cite{ootomo2024cagra,zhang2026distribution}.
At billion scale, proximity graphs also incur substantial storage overhead,
while updates and full rebuilds remain expensive~\cite{singh2021freshdiskann,manohar2024parlayann}.

These limitations motivate revisiting the Inverted File (IVF) index as a
simpler alternative~\cite{jegou2011searching,johnson2019billion,baranchuk2018revisiting}.
IVF partitions vectors with $k$-means and searches selected lists through coarse
centroid routing and fine scanning~\cite{lloyd1982least,jegou2011searching}.
Its flat structure has lower auxiliary overhead, more regular memory access,
and cheaper construction and rebuilding than graph indexes. IVF is also
naturally batch-friendly: dense centroid routing maps well to GPUs, while list
scanning exploits contiguous storage~\cite{johnson2019billion,baranchuk2018revisiting,faiss2024}.
These properties are attractive for vector databases that prioritize data
freshness, predictable cost, and simple deployment~\cite{wang2021milvus,sun2025gaussdb}.

\begin{figure*}
    \centering
    \includegraphics[width=\textwidth]{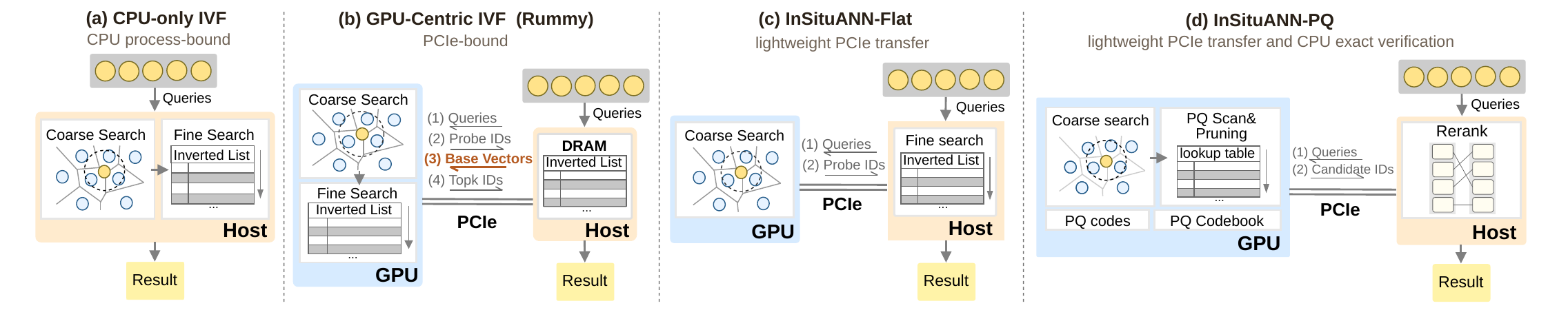}
    \caption{Overview of the four vector search methods.}
    \label{fig:intro4method}
\end{figure*}

Modern systems accelerate IVF by offloading its regular, data-parallel distance
computations to GPUs~\cite{johnson2019billion,zhang2024fast}. Full-vector
billion-scale indexes, however, do not fit on one commodity GPU:
SIFT-1B alone occupies 512\,GB in \texttt{float32} or 128\,GB in 8-bit
storage, excluding auxiliary index data~\cite{jegou2011searching,simhadri2024results}.
A practical design therefore keeps base vectors in host DRAM and uses the GPU
for compute-intensive stages~\cite{zhang2024fast,peng2026svfusion,karthik2025bang}.
However, in GPU-centric systems such as Rummy
(Figure~\ref{fig:intro4method}(b)), GPU fine search requires candidate vectors
or pages to cross PCIe during query processing~\cite{zhang2024fast}. The amount
of transferred data grows with the probed set, even though each distance
evaluation performs little computation~\cite{williams2009roofline,li2025scaling}. Fetching only 2\% of
SIFT-1B moves 2.56~GB and takes at least 80~ms at the theoretical peak bandwidth
of PCIe 4.0 x16. Bulk base-vector movement can therefore dominate GPU-side fine
search at billion scale.

These observations suggest a stage-aware placement rule: keep compact, regular
computation on the GPU, but execute bandwidth-bound vector scans where the base
vectors reside~\cite{williams2009roofline}. In IVF, this rule yields a natural
split between dense GPU centroid routing and CPU fine search over host-resident
vectors. Based on this split, we propose \textbf{\ourname{}}, a heterogeneous
IVF engine with two hardware-matched components
(Figure~\ref{fig:intro4method}(c)):

\begin{itemize}[leftmargin=1em,topsep=0.2em,itemsep=0.25em,parsep=0pt]
\item \textbf{GPU coarse search:} Dense query--centroid routing runs as batched,
Tensor Core-accelerated matrix multiplication, exploiting its regular
parallelism and high arithmetic intensity.
\item \textbf{CPU in-situ fine search:} The CPU scans selected contiguous lists
directly in host DRAM. Query tiling and multi-core SIMD computation exploit
local memory bandwidth without transferring base-vector payloads over PCIe.
\end{itemize}
The two-component design first yields the \ourname{}-Flat path. After GPU coarse
search, the CPU scans selected host-resident vectors and computes exact
distances for ranking. This eliminates query-time base-vector transfers;
however, scanning uncompressed vectors can still be limited by host-memory
bandwidth.

\ourname{}-PQ reduces this cost by using Product Quantization (PQ) as a GPU
candidate-reduction layer (Figure~\ref{fig:intro4method}(d))~\cite{jegou2010product}.
The GPU scans compact PQ codes and returns a bounded set of candidate IDs over
PCIe; the CPU then computes exact distances over the corresponding original
vectors. PQ therefore changes candidate selection, not final-ranking semantics.
The codes for 1.4B vectors require at most 28~GB of VRAM, enabling GPU filtering
while full vectors remain in host DRAM.

This paper makes the following contributions:

\begin{itemize}[leftmargin=1em]

    \item \textbf{Shattering the PCIe Bottleneck for Billion-Scale IVF.}
    We identify PCIe movement of base vectors as the fundamental bottleneck in
    GPU-centric IVF once they exceed GPU memory. This motivates a
    hardware-matched architecture: dense routing runs on the GPU, while the CPU
    scans host-resident vectors in situ. By
    eliminating query-time base-vector transfers, \ourname{} achieves
    $104.9\times$--$4298.2\times$ higher throughput than Rummy~\cite{zhang2024fast}.

    \item \textbf{Bottleneck-Driven Cross-Device IVF Optimization.}
    Guided by stage-level profiling, we optimize each component at its measured
    bottleneck: GPU residual-PQ pruning bounds the verification workload, CPU
    SIMD kernels verify candidates against exact host-resident vectors, and
    pipeline overlap reduces idle time across routing, pruning, and fine search.
    The resulting Flat and PQ modes
    provide more than an order-of-magnitude speedup over CPU-only IVF while
    retaining IVF's simple, update-friendly layout.

    \item \textbf{Graph-Outperforming Search and Ultra-Fast Construction.}
    At matched Recall@10 on billion-scale datasets, \ourname{} achieves
    $2.4\times$--$4.6\times$ higher QPS than DiskANN~\cite{jayaram2019diskann}.
    Its graph-assisted builder constructs a CAGRA
    graph~\cite{ootomo2024cagra} only over IVF centroids to accelerate centroid
    search during training and full-dataset assignment. It completes SIFT-1B in
    \textbf{5.2 minutes}, compared with
    \textbf{30.4 hours} for HNSW, without materializing a billion-node
    base-vector graph.

\end{itemize}

\section{Background}\label{sec:background}

\subsection{ANN Search and IVF}
\label{sec:bg:ivf}

Given a database $\mathcal{D}=\{x_1,\ldots,x_N\}$ of $N$ vectors in
$\mathbb{R}^d$ and a query vector $q$, exact $K$-nearest neighbor search
returns the $K$ vectors in $\mathcal{D}$ with the smallest distances to
$q$. Approximate nearest neighbor search (ANNS) accelerates this process
by building an index over $\mathcal{D}$ and returning an approximate
result set, typically trading a small loss in recall for much higher
throughput and lower latency~\cite{arya1993approximate,jegou2011searching,malkov2018efficient}.

The inverted file (IVF) index is one of the most widely used ANNS
structures due to its simple layout and hardware-friendly execution
pattern~\cite{jegou2011searching,babenko2014inverted,johnson2019billion}. During index construction, IVF partitions the database into
$N_{\mathrm{list}}$ clusters. Let
$\mathcal{C}=\{c_1,\ldots,c_{N_{\mathrm{list}}}\}$ be the centroid set. Each vector in the database
 is assigned to its nearest centroid and stored in the corresponding
inverted list. In practical implementations, vectors in the same list are
often laid out contiguously, so that examining a list translates into a
sequential memory scan.

At query time, IVF decomposes search into two stages. The first stage,
\emph{coarse search}, compares the query with all centroids and selects
the top-$P$ closest clusters, where $P$ is the probe count:
\begin{equation}
    \mathcal{I}_q =
    \operatorname*{arg\,min}_{\substack{\mathcal{I}\subseteq\{1,\ldots,N_{\mathrm{list}}\}\\|\mathcal{I}|=P}}
    \sum_{i\in\mathcal{I}} \mathrm{dist}(q,c_i).
\end{equation}
For a batch of queries, this stage becomes a dense query--centroid
distance computation, which exposes regular parallelism and maps well to
GPU execution~\cite{johnson2019billion}.

The second stage, \emph{fine search}, scans the inverted lists selected
by coarse search and returns the approximate top-$K$ results:
\begin{equation}
    \widehat{\mathcal{N}}_K(q)
    =
    \operatorname*{arg\,min}_{\substack{
        \mathcal{S} \subseteq \bigcup_{i \in \mathcal{I}_q} \mathcal{L}_i \\
        |\mathcal{S}| = K
    }}
    \sum_{x \in \mathcal{S}} \mathrm{dist}(q,x).
\end{equation}
where $\mathcal{L}_i$ denotes the list associated with centroid $c_i$.
Unlike coarse search, fine search is dominated by moving candidate-vector
data through the memory hierarchy~\cite{williams2009roofline,zhang2024fast,li2025scaling}. This distinction is central to
\ourname{}: IVF naturally separates a dense routing stage from a
data-intensive candidate-evaluation stage, creating an opportunity for
cross-device placement.
\subsection{Product Quantization for IVF Search}
\label{sec:bg:pq}

IVF reduces search cost by probing only a subset of inverted lists, but
the fine-search stage may still examine a large number of candidates at
billion scale. Scanning full-precision vectors for all candidates creates
substantial memory traffic. Product Quantization (PQ) is therefore
widely used to represent candidate vectors with compact codes and to
estimate distances before exact verification~\cite{jegou2010product}.

PQ partitions a $d$-dimensional vector into $\mathrm{PQ}_M$ disjoint sub-vectors and
quantizes each subspace independently. For a vector
$x=[x^{(1)},\ldots,x^{(\mathrm{PQ}_M)}]$, where
$x^{(m)}\in\mathbb{R}^{d/\mathrm{PQ}_M}$, PQ
learns one codebook
$\mathcal{C}^{(m)}=\{c^{(m)}_1,\ldots,c^{(m)}_K\}$ per subspace and
encodes each sub-vector by its nearest codeword:
\begin{equation}
    \mathrm{PQ}(x) = (i_1,\ldots,i_{\mathrm{PQ}_M}), \qquad
    i_m = \arg\min_{j \in \{1,\ldots,K\}}
          \mathrm{dist}\!\left(x^{(m)}, c^{(m)}_j\right).
\end{equation}
When $K=256$, each sub-code is stored in one byte, so a PQ code occupies
$\mathrm{PQ}_M$ bytes. 

At query time, PQ distances are commonly evaluated with Asymmetric
Distance Computation (ADC)~\cite{jegou2010product}. Given a query
$q$, ADC first builds a lookup table:
\begin{equation}
    \mathrm{LUT}[m,j] =
    \mathrm{dist}\!\left(q^{(m)}, c^{(m)}_j\right),
\end{equation}
and then estimates the distance to an encoded vector by summing the
lookup entries selected by its PQ code:
\begin{equation}
    \widetilde{\mathrm{dist}}(q,x) =
    \sum_{m=1}^{\mathrm{PQ}_M} \mathrm{LUT}[m,i_m].
\end{equation}
Thus, candidate evaluation can be reduced to compact-code scans and
lookup-table accumulations~\cite{andre2017accelerated,andre2019quicker}. In \ourname{}, PQ is used as a pruning layer:
approximate ADC scores select a bounded candidate set, and the final
ranking is still computed by exact distances over the retained original
vectors.

\subsection{Bandwidth Mismatch in Hybrid IVF Search}
\label{sec:bg:bottleneck}

Hybrid CPU--GPU IVF systems expose a bandwidth mismatch across memory paths~\cite{zhang2024fast,tian2025towards}.
Coarse search uses a compact centroid table that can reside in GPU memory,
whereas fine search must access host-resident base vectors from the selected
IVF lists. These vectors can either be scanned in situ by the CPU from host
DRAM, or transferred over PCIe for GPU-side evaluation in GPU-centric pipelines~\cite{zhang2024fast}.
Because PCIe bandwidth is much lower than both host-DRAM bandwidth and GPU
local-memory bandwidth, moving large volumes of candidate vectors can become
the dominant cost in fine search~\cite{williams2009roofline,li2025scaling}.
\begin{table}
\centering
\small
\caption{Bandwidth gap for transferring 2.56~GB of candidate vectors.}
\label{tab:bandwidth-mismatch}
\resizebox{0.95\columnwidth}{!}{
\begin{tabular}{lcc}
\toprule
\textbf{Data path} & \textbf{Bandwidth scale} &
\textbf{Time for 2.56~GB} \\
\midrule
GPU local memory & TB/s scale & $\sim$1--3 ms \\
Host DDR5 DRAM & hundreds of GB/s & several ms \\
PCIe 5.0 x16 & $\leq$64 GB/s & $\geq$40.0 ms \\
PCIe 4.0 x16 & $\leq$31.5 GB/s & $\geq$81.3 ms \\
\bottomrule
\end{tabular}}
\end{table}

Table~\ref{tab:bandwidth-mismatch} shows the resulting mismatch.  Recent GPUs
provide TB/s-scale local memory bandwidth and high arithmetic throughput, but
host--device PCIe bandwidth remains only tens of GB/s.  For SIFT-1B, the raw
base vectors occupy 128 GB.  Fetching just 2\% of  base vectors moves 2.56~GB over PCIe, taking at least 40~ms on PCIe 5.0 x16 and 80~ms on PCIe 4.0 x16, while in-situ CPU scan avoids this host--device transfer.

This motivates a stage-aware placement strategy: keep dense and reusable
routing data on the GPU, but execute large host-resident vector scans close to
the data instead of moving the vectors across PCIe.

\section{\ourname{}-Flat Overview}
\label{sec:overview}

\begin{figure}
  \centering
  \vspace{-0.5em}
  \includegraphics[width=\columnwidth, trim=0 0 0 0, clip]{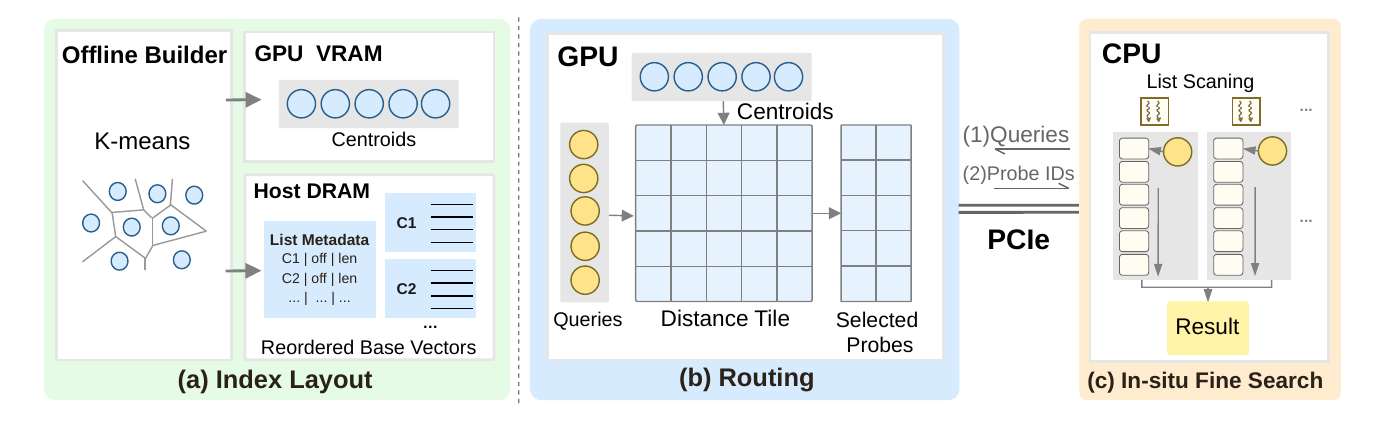}
  \caption{\ourname{}-Flat architecture.}
  \vspace{-0.6em}
  \label{fig:overview}
\end{figure}

\subsection{System Architecture}
\label{sec:overview:arch}
Figure~\ref{fig:overview} illustrates the architecture of
\ourname{}-Flat.

\textbf{Index layout.}
The offline builder trains IVF centroids, assigns each base vector to its
nearest centroid, and stores the reordered vectors in cluster-major order.
Each IVF list is represented by compact metadata that records its offset,
length, and vector identifiers. This layout allows \ourname{}-Flat to
translate probed cluster identifiers into contiguous host-memory ranges during
search.

\textbf{GPU routing.}
The centroid matrix is kept resident in GPU memory. For each query batch, the
GPU computes query--centroid distances and selects the closest IVF lists for
each query. Only the selected list identifiers and compact routing metadata are
returned to the host.

\textbf{Host in-situ fine search.}
The original base vectors remain in host DRAM throughout query processing.
Given the probed IVF lists, the CPU scans the corresponding contiguous
host-memory ranges and computes exact distances over the original vectors.
Thus, fine search is performed in situ, without transferring base-vector
payloads to the GPU.

\textbf{Pipeline scheduler.}
\ourname{}-Flat pipelines GPU routing and host-side fine search across query
batches. While the GPU routes one batch, the CPU scans and verifies candidates
for another, reducing idle time and improving end-to-end throughput.

\subsection{Query Processing Workflow}
\label{sec:overview:query}

\begin{itemize}[leftmargin=1em]
  \item \textbf{Batching and GPU routing.}
  Incoming queries are grouped into batches and copied to the GPU. The GPU
  compares the batch against the resident centroid matrix and selects the
  closest IVF lists for each query.

  \item \textbf{Probe-list lookup.}
  The host uses the returned list identifiers to look up the corresponding IVF
  list offsets and lengths. Because the index is stored in cluster-major order,
  each probed list maps to a contiguous host-memory range.

  \item \textbf{CPU in-situ fine search.}
  The CPU scans the selected host-resident lists and computes exact distances
  to the original base vectors. During the scan, each worker maintains a local
  top-$K$ set; after all workers finish, their local sets are merged into the
  final top-$K$ neighbors for each query.
\end{itemize}

Thus, query-time PCIe traffic is limited to queries and compact routing
metadata, while base vectors remain in host memory.

\subsection{Fine-Search Placement}
\label{sec:access-locality}

The CPU in-situ placement avoids moving original vectors during fine search.
The access-concentration analysis in Appendix~\ref{app:access-locality} finds no
small hot set that captures most fine-search accesses, so caching cannot be
assumed to eliminate PCIe traffic.

\textbf{Comparison paths.}
We compare three backends under the same IVF routing and candidate set:

\noindent\hspace*{2em}CPU in-situ: keeps the selected IVF lists in host memory and computes
exact distances on the CPU.\par
\noindent\hspace*{2em}GPU uncached: transfers the selected lists to the GPU and performs
exact fine search there.\par
\noindent\hspace*{2em}GPU cache: adds a static, list-granular GPU cache to GPU uncached,
with uncached lists following the same transfer path.\par

Cache contents are selected on a calibration query set and evaluated on a
disjoint held-out set. We evaluate all datasets across multiple cache
capacities, recall targets, and batch sizes.

\begin{figure*}[t]
    \centering
    \includegraphics[width=0.68\textwidth]{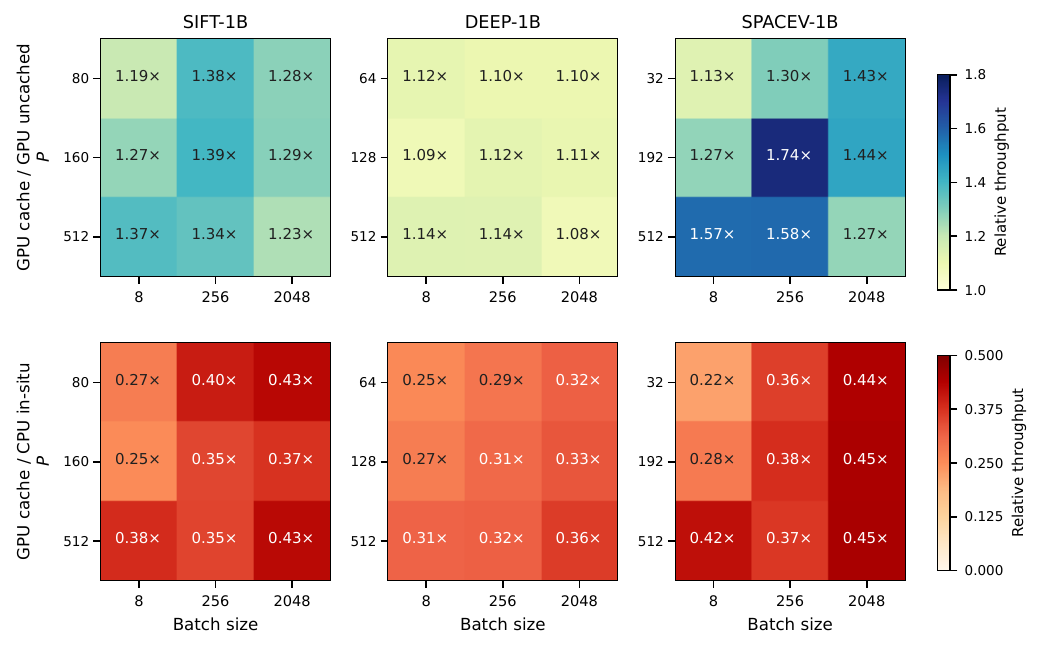}
    \caption{GPU-cache throughput across recall targets and batch sizes.}
    \label{fig:gpu-cache-full-matrix-main}
\end{figure*}

\textbf{Results and implication.}
At Recall@10 $\approx 0.90$ and batch size 2048, a 24~GiB GPU cache improves
GPU uncached throughput by $1.10\times$--$1.43\times$ across the three
datasets. Even at its best, GPU cache remains below CPU in-situ throughput,
so caching narrows but does not close the placement gap
(Figure~\ref{fig:gpu-cache-full-matrix-main}).

\section{Optimizations}
\label{sec:opt}

After \ourname{}-Flat removes query-time base-vector transfers over PCIe, the
remaining costs are GPU coarse routing, CPU in-situ fine search over
host-resident vectors, and coordination between GPU and CPU work. We optimize
the search path as a sequence of bounded stages, keeping each stage's work
explicit and controllable.

\begin{figure}
  \centering
  \includegraphics[width=\columnwidth, trim=0 0 0 0, clip]{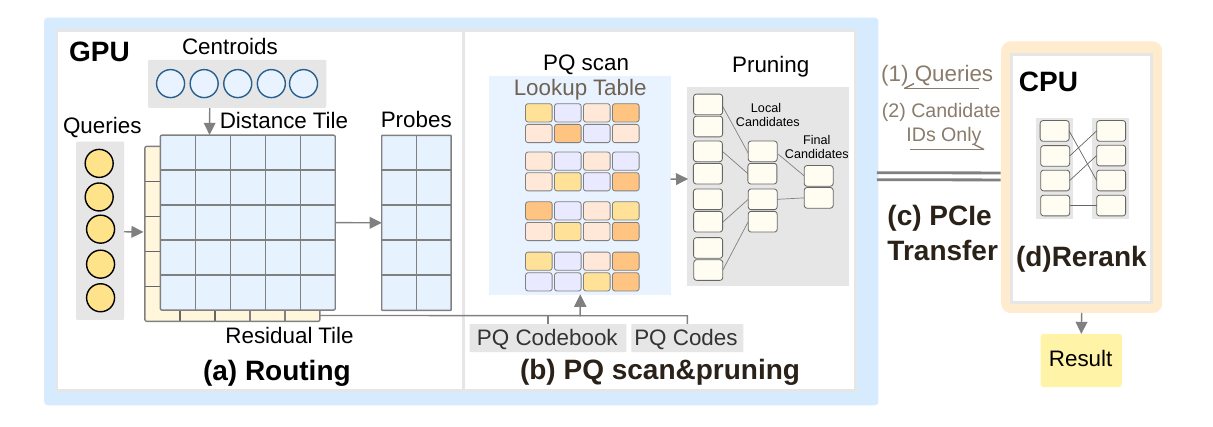}
  \caption{\ourname{}-PQ Architecture.}
  \vspace{-0.7em}
  \label{fig:gpu-pq-over}
\end{figure}

\ourname{}-PQ adds residual-PQ candidate pruning before exact
evaluation: as shown in Figure~\ref{fig:gpu-pq-over}, by default, PQ pruning is offloaded to the GPU over compact PQ
codes, and only a bounded set of candidate IDs is returned for exact CPU
distance computation. Notably, final ranking is computed using exact
distances over the retained original vectors; PQ is used only for candidate
reduction~\cite{jegou2010product}. We additionally implement
a CPU-only PQ pruning mode for ablation studies and as a fallback when GPU PQ
offloading is unavailable.

\subsection{GPU coarse search}
\label{sec:opt:coarse}

The first stage of IVF search identifies the nearest IVF lists for each query.
\ourname{} keeps this stage on the GPU because query--centroid distance
computation is dense, regular, and operates over a compact centroid matrix.

\textbf{Dense centroid evaluation.}
Let $Q$ be a batch of queries and $C$ be the centroid matrix. \ourname{}
computes query--centroid scores as a batched dense operation. For L2 search,
distances are derived from inner products and precomputed norms:
\begin{equation}
    \|q-c\|_2^2 = \|q\|_2^2 + \|c\|_2^2 - 2q^\top c .
\end{equation}
This turns query--centroid distance computation into a batched dense matrix
operation, which maps efficiently to GPUs and amortizes kernel-launch and
host--device transfer overhead across a query~\cite{johnson2019billion}.

\textbf{On-device list selection.}
After computing query--centroid distances, \ourname{} selects the nearest
$P$ IVF lists for each query directly on the GPU. This GPU routing stage is
shared by both \ourname{}-Flat and \ourname{}-PQ; the selected list identifiers
are later used either for host-side full-vector scanning or for PQ-code
pruning.

Rather than transferring the full $B \times N_{\mathrm{list}}$ distance matrix
to the host, the GPU performs row-wise top-$P$ selection on device and returns
only the selected list identifiers and optional distances, whose size is
$O(BP)$. At billion scale, this reduction is substantial: the total number of
IVF lists is on the order of hundreds of thousands, while $P$ is orders of
magnitude smaller. \ourname{} implements this selection with tiled GPU kernels
that compute tile-local top-$P$ candidates and merge them on device. The host
then maps the selected list identifiers to contiguous list ranges for the next
fine-search stage.

\textbf{Finer IVF partitioning.}
At billion scale, \ourname{} uses a larger $N_{\mathrm{list}}$ to create finer IVF partitions. More fine-grained cells shorten each inverted list and make the probed lists more localized, reducing the number of host-resident base vectors scanned during fine search~\cite{jegou2011searching,baranchuk2018revisiting}. This increases coarse-routing work, but the cost is controllable because routing is a dense GPU operation over a compact centroid table. In the pipelined execution, GPU routing and CPU in-situ fine search run on different batches concurrently, so the additional routing work can be absorbed more smoothly and the pipeline remains better balanced across stages.
This improves the recall--throughput trade-off up to the point where routing cost begins to dominate. Appendix~\ref{app:nlist-sensitivity} reports $N_{\mathrm{list}}$ sensitivity.

\subsection{CPU Exact Distance Evaluation}
\label{sec:opt:cpu}

Exact distance evaluation is the common final stage of both \ourname{}-Flat
and \ourname{}-PQ. It computes exact distances over original vectors that remain
in host memory. \ourname{} optimizes this stage for locality, reuse, and
parallel scan efficiency.

\textbf{List-major contiguous scans.}
The builder stores base vectors in list-major order, where each IVF list
corresponds to a contiguous host-memory range. Once the GPU returns selected
IVF-list identifiers, the CPU maps them to offsets and lengths and scans the
corresponding ranges sequentially. This layout turns fine search into streaming
reads over host-resident lists rather than scattered accesses over the original
vector order.

\textbf{Exact distance computation.}
\ourname{} uses a shared exact distance verifier as the final-ranking stage.
The verifier uses type-specialized SIMD kernels. For floating-point vectors,
it uses AVX-512/FMA vectorized L2 kernels; for byte-stored vectors, it uses
AVX-512BW widening and \texttt{\_mm512\_madd\_epi16}-style accumulation to
compute integer distances efficiently~\cite{intel_avx_manual}. Each worker
maintains a thread-local top-$K$ heap during the scan, and the per-thread
results are merged into the final per-query top-$K$ output.

\textbf{Touched-list scheduling and query tiling.}
For each query batch, \ourname{}-Flat organizes CPU fine search by touched IVF
list. If multiple queries select the same list, the list is streamed from host
memory only once and evaluated against those queries together. This avoids
repeatedly reading the same host-memory range for different queries. During the
scan, the default kernel compares each loaded base vector with up to four
queries before loading the next vector, reusing the vector load and amortizing
loop overhead across queries.

\subsection{Residual-PQ Candidate Reduction}
\label{sec:opt:pq}

Exact verification can dominate latency when the probed IVF lists produce a
large candidate volume. \ourname{}-PQ adds a residual-PQ candidate-reduction stage before exact distance evaluation~\cite{jegou2010product,kalantidis2014locally}, scanning compact residual codes, keeping a bounded set of promising candidates, and passing only those candidates to the exact verifier.

\textbf{Residual encoding.}
For each base vector $x$, \ourname{}-PQ encodes the residual $x-c(x)$, where
$c(x)$ is the IVF centroid assigned to $x$. Compared with global PQ that
quantizes the original vector space directly, residual PQ quantizes only the
within-list offset after coarse assignment. These residuals have smaller
dynamic range and lower variance within each IVF list, leading to more accurate
distance estimates under the same code budget~\cite{kalantidis2014locally}.
Residual encoding therefore separates coarse routing from fine-grained
candidate ordering: the centroid determines the searched IVF list, while the
residual code approximates the vector's local offset inside that list.
Residual codes are stored in the same list-major order as the original vectors,
so each probed list corresponds to contiguous code and vector ranges.

\textbf{Approximate reduction, exact ranking.}
For each probed IVF list with centroid $c$, \ourname{}-PQ forms the query residual
$r_q=q-c$ and builds a lookup table
\[
    \mathrm{LUT}_m[j] = \|r_q^{(m)} - p_m[j]\|_2^2 ,
\]
where $p_m[j]$ is the $j$-th PQ codeword in subspace $m$. For a candidate
$x$ with residual-PQ code $(a_1(x),\ldots,a_{\mathrm{PQ}_M}(x))$, its approximate distance is
computed by table lookup and accumulation:
\[
    \tilde{d}(q,x) = \sum_{m=1}^{\mathrm{PQ}_M} \mathrm{LUT}_m[a_m(x)] .
\]
\ourname{}-PQ retains the candidates with the smallest $\tilde{d}(q,x)$ and maps
their identifiers back to the original host-resident vectors. Final ranking is
computed using exact distances over these original vectors.

\subsection{Compact PQ-Code Offloading}
\label{sec:opt:pqgpu}

\textbf{GPU-side PQ-code offloading.}
\ourname{}-PQ offloads residual-PQ pruning to the GPU by keeping compact
residual-PQ codes in otherwise idle GPU memory. Since IVF routing state occupies
only hundreds of MB, most VRAM remains unused when the GPU is used only for
coarse search. We therefore offload PQ codes, rather than full base vectors, on
the GPU and scan them there for candidate pruning. The original vectors remain
in host memory and are accessed only by the CPU for exact final ranking.

\textbf{GPU-side task setup.}
\ourname{}-PQ avoids building per-batch fine-search metadata on the CPU.
Given the probed IVF-list identifiers and list lengths, the GPU constructs the
chunk and group descriptors needed for PQ scanning and merging, using prefix
scans to assign output offsets. It also reuses preallocated GPU
workspaces for queries, descriptors, intermediate candidates, and merge outputs, removing repeated CPU-side metadata construction and per-batch GPU memory
allocation.

\textbf{Chunked PQ scan with local top-$K$.}
After probe-list lookup, each selected IVF list is mapped to a contiguous
residual-PQ code range. Since IVF lists can vary substantially in length,
\ourname{}-PQ splits each probed list into fixed-size chunks and treats each
chunk as an independent GPU work unit. This makes GPU work more balanced and
regular. Within each chunk, the GPU scans residual-PQ codes and computes ADC-style
approximate distances using GPU-resident codebooks and per-query lookup tables.
Rather than materializing distances for all candidates, each chunk retains only
a small local top-$K$ candidate set for later merging. This bounds intermediate
output early and reduces global-memory traffic in the pruning pipeline.

\textbf{Hierarchical candidate merging.}
\ourname{}-PQ reduces candidates through a chunk--group hierarchy. A chunk is a
fixed-size segment of a probed IVF list and is the basic GPU scan unit; a group
contains multiple chunks and is the first merge unit. Chunk-local candidates are
merged into group-level candidates, which are then merged into a bounded
per-query handoff set.

The handoff budget $N_{\mathrm{rerank}}$ is tunable and controls how many candidates are sent to
exact CPU re-ranking. For a batch of size $B$, device-to-host traffic is
therefore bounded by $O(BN_{\mathrm{rerank}})$, instead of scaling with the number of probed
vectors.

\textbf{Fused-first candidate generation.}
\ourname{}-PQ fuses chunk scanning with the first group-level merge. Instead of
writing chunk-local candidates to global memory and launching a separate kernel
to merge them, one GPU block processes a group of chunks and directly emits the
merged group-level candidate set. The block keeps local candidates in
registers/shared memory during the scan, eliminating the chunk-level
intermediate write/read round trip. This reduces global-memory traffic and
removes one kernel stage from the pruning pipeline.

\subsection{Pipeline}
\label{sec:opt:pipeline}

\ourname{} executes query processing as a staged CPU--GPU pipeline across
batches. The GPU handles coarse search and, in \ourname{}-PQ, residual-PQ pruning,
while the CPU performs exact distance evaluation over host-resident original
vectors. Running these stages serially would underutilize one device while the
other is active, so \ourname{} overlaps GPU work for one batch with CPU
re-ranking for another.

\textbf{Overlapping GPU and CPU work.}
\ourname{} uses a one-batch-deep producer--consumer pipeline to overlap GPU and
CPU execution across batches. The main thread performs GPU routing and, in
\ourname{}-PQ, GPU-side pruning for batch $i+1$, while a CPU worker performs
exact re-ranking for batch $i$. The pipeline improves steady-state
throughput.

\textbf{Compact handoff.}
The stage interfaces remain compact throughout the pipeline. Routing emits only
$O(BP)$ probe metadata for batch size $B$ and probe count $P$, while
\ourname{}-PQ emits at most $O(BN_{\mathrm{rerank}})$ candidate identifiers, where
$N_{\mathrm{rerank}}$ is the
exact re-ranking budget. Exact re-ranking reads the corresponding original
vectors directly from host memory. Thus, PCIe traffic scales with probe metadata
and retained candidate IDs, rather than with all vectors in the probed IVF
lists, enabling very compact handoff.

\section{Evaluation}\label{sec:eval}
\textbf{Hardware environment.}
Our primary server has one NVIDIA RTX 5090 GPU with 31.8 GB VRAM, two Intel
Xeon Platinum 8469C CPUs with 96 physical cores in total, and 1032 GB of DRAM.
It runs Ubuntu 24.04 LTS and CUDA 13.0. Full vectors remain host resident, and
we do not maintain extra serving-time copies of the base-vector payload.
H100 results appear in Appendix~\ref{app:hardware-portability}.

\textbf{Datasets.}
We evaluate SIFT~\cite{jegou2011searching}, DEEP~\cite{DEEP}, and SPACEV~\cite{SPACEV} at 100M and billion scales.
Table~\ref{tab:datasets} reports the host-resident full-vector payload and the
compact residual-code payload used by \ourname{}.  Full vectors 
remain in host DRAM, while the GPU stores the compact residual-code payload
together with small codebooks, routing/list metadata, and runtime buffers.  We
use 10K queries for each dataset. PQ uses $K=256$ 8-bit 
subcodes, fp32 LUTs, and an exact re-ranking budget $N_{\mathrm{rerank}}=512$ after PQ pruning. The residual-PQ
codebooks are below 0.2 MB and are therefore omitted from the table.
\begin{table}
\centering
\caption{Dataset sizes and serving-time memory footprint.}
\label{tab:datasets}
\scriptsize
\setlength{\tabcolsep}{3pt}
\renewcommand{\arraystretch}{1.1}
\begin{tabular}{@{}lcccccc@{}}

\toprule
Dataset & Dim & Base & Type & Host Vec. & $\mathrm{PQ}_M$ & PQ Codes \\
\midrule
DEEP-1B & 96 & 1B & float32 & 384~GB & 16 & 16~GB \\
SPACEV-1B & 100 & 1.4B & int8 & 140~GB & 20 & 28~GB \\
SIFT-1B & 128 & 1B & uint8 & 128~GB & 16 & 16~GB \\
\midrule
DEEP-100M & 96 & 100M & float32 & 38.4~GB & 32 & 3.2~GB \\
SPACEV-100M & 100 & 100M & int8 & 10.0~GB & 50 & 5.0~GB \\
SIFT-100M & 128 & 100M & uint8 & 12.8~GB & 32 & 3.2~GB \\
\bottomrule
\end{tabular}
\vspace{-0.9em}
\end{table}

\textbf{\ourname{} configuration.}
To make comparisons controlled, all \ourname{} variants share the same IVF
partitioning. We use $N_{\mathrm{list}}=0.5\mathrm{M}$ at billion scale
and $N_{\mathrm{list}}=0.125\mathrm{M}$ at 100M scale, where
$1\mathrm{M}=2^{20}$. We sweep $P$ over a discrete grid
spanning 1 to 512 and evaluate four batch sizes: 8, 32, 256, and 2048.
\ourname{}-Flat uses 64 CPU threads for fine search, while
\ourname{}-PQ uses 24 CPU threads for the bounded exact re-ranking stage.

\textbf{Baselines.}
We compare \ourname{} with Faiss-CPU~\cite{faiss2024},
Rummy~\cite{zhang2024fast}, DiskANN~\cite{jayaram2019diskann}, and
PilotANN~\cite{gui2026pilotann}. We additionally evaluate
PipeANN~\cite{guo2025pipeann} as a
graph-based baseline on SIFT-1B and SPACEV-1B. Faiss-CPU is configured with 64 OpenMP
threads and is evaluated with the same batch sizes and query sets as
\ourname{}. Faiss-CPU-IVFPQ uses the same $N_{\mathrm{list}}$ and the same PQ code size as
\ourname{}-PQ when applicable, and is evaluated with 64 OpenMP threads over
a matched $P$ sweep. Rummy is run as a single-GPU
heterogeneous IVF baseline. 

DiskANN is evaluated as an on-disk Vamana baseline with a patched batched search
runner in SSD-based serving mode. We evaluate HNSW~\cite{malkov2018efficient}
with 64 CPU threads and the same batch sizes as \ourname{}, using the largest
configurations that complete within available memory. On SPACEV-1B, only
$M_{\mathrm{HNSW}}=10$ and $ef_{\mathrm{construction}}=32$ completes among
the evaluated settings, illustrating its billion-scale construction limit.
Full baseline settings appear in Appendix~\ref{app:evaluation-details}.

We also considered two recent GPU-based ANNS systems,
FusionANNS~\cite{tian2025towards} and BANG~\cite{karthik2025bang}.
FusionANNS could not be evaluated because its implementation was not
publicly available at the time of evaluation. We evaluated BANG using
its released implementation but were unable to reproduce the reported
performance on our hardware. Accordingly,
neither system is included in the quantitative comparison.

\subsection{End-to-End Performance}
\label{subsec:main_results}
\begin{figure*}
    \centering
    \includegraphics[width=0.98\textwidth]{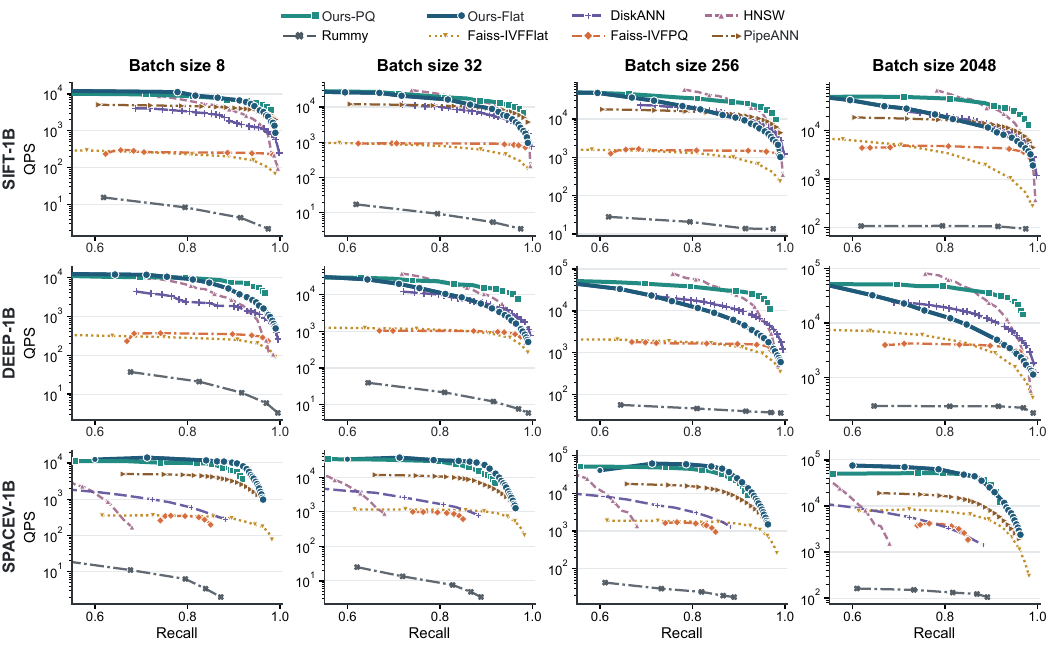}
    \caption{
        End-to-end QPS--Recall trade-offs on SIFT-1B, DEEP-1B, and SPACEV-1B across batch sizes 8, 32, 256, and 2048.
    }
    \vspace{-0.6em}
    \label{fig:graph_compare_1b}

\end{figure*}

Figure~\ref{fig:graph_compare_1b} reports the end-to-end QPS--Recall trade-offs
on the three billion-scale datasets. Appendix~\ref{app:results-100m} reports
100M-scale results.

At billion scale, \ourname{} provides strong
QPS--Recall frontiers across the three billion-scale datasets. \ourname{}-PQ
delivers the highest throughput over most practical recall targets by moving
candidate reduction to the GPU and leaving only bounded exact verification on
the CPU. \ourname{}-Flat complements it in the highest-recall region, where
retaining more candidates for exact ranking is more important than 
PQ pruning.

The largest gap is against Rummy. At comparable recall, \ourname{}-PQ improves
throughput by $271.9\times$--$4298.2\times$ for $\mathrm{bs}=8,32$ and by
$104.9\times$--$1819.7\times$ for $\mathrm{bs}=256,2048$. This follows from
their different data-movement strategies: Rummy transfers fine-search vectors
over PCIe for GPU evaluation, whereas \ourname{} scans host-resident vectors
in situ and avoids this bulk movement. Section~\ref{subsec:cdf_memory}
reports the corresponding latency distributions.

Against Faiss-CPU-IVFPQ, \ourname{}-PQ improves throughput by
$8.6\times$--$47.8\times$ for $\mathrm{bs}=8,32$ and
$4.1\times$--$36.9\times$ for $\mathrm{bs}=256,2048$. Relative to
Faiss-CPU-IVFFlat, \ourname{}-Flat improves throughput by
$2.0\times$--$39.9\times$ for $\mathrm{bs}=8,32$ and
$1.7\times$--$34.6\times$ for $\mathrm{bs}=256,2048$.

Batch size and dataset geometry jointly shape the two paths. On SIFT-1B and
DEEP-1B, residual-PQ pruning is highly effective, so \ourname{}-PQ leads the
high-throughput frontier across most medium- and high-recall targets; its
advantage becomes more pronounced at larger batches, especially above
Recall@10 $0.9$. \ourname{}-Flat remains competitive at smaller batches and
lower recall, and extends the frontier at the highest-recall end by avoiding
early candidate pruning. SPACEV-1B is less favorable to PQ: candidate quality
degrades earlier as recall rises, while Flat remains competitive over a wider
range and reaches the highest recall points. The two modes are therefore
complementary: PQ maximizes throughput when pruning is reliable, whereas Flat
is preferable when recall or robustness is the priority.
    
Against DiskANN on SIFT-1B, \ourname{}-PQ improves throughput by
$2.4\times$--$4.3\times$ at Recall@10 $\geq 0.90$ and
$2.6\times$--$4.6\times$ at Recall@10 $\geq 0.95$ across batch sizes 8--2048.
On DEEP-1B, the corresponding improvements are $2.7\times$--$4.0\times$ and
$3.2\times$--$4.6\times$, respectively.
Above Recall@10 $0.9$, \ourname{} reaches up to $14.9\times$ and
$16.2\times$ higher throughput than HNSW on SIFT-1B and DEEP-1B,
respectively. SPACEV-1B further exposes the construction limit: the largest
memory-fitting HNSW configuration uses the reduced setting reported in
Appendix~\ref{app:evaluation-details} and remains below Recall@10 $0.7$, while
\ourname{} reaches substantially higher recall.

On the two datasets with PipeANN results, \ourname{}-PQ improves throughput at
comparable recall on both SIFT-1B and SPACEV-1B, reaching up to $3.55\times$
higher throughput.

\subsection{Latency Distribution}
\label{subsec:cdf_memory}

Throughput alone does not characterize whether an ANNS system provides
predictable response times. We therefore measure batch latency at
Recall@10$\approx$0.9 using only full-size steady-state batches. Within each
repeat, we exclude the first two full batches and the final full batch to remove
pipeline fill and drain effects.

\begin{figure}
    \centering
    \includegraphics[
        width=0.28\textwidth,
        trim=0 6 0 8,
        clip
    ]{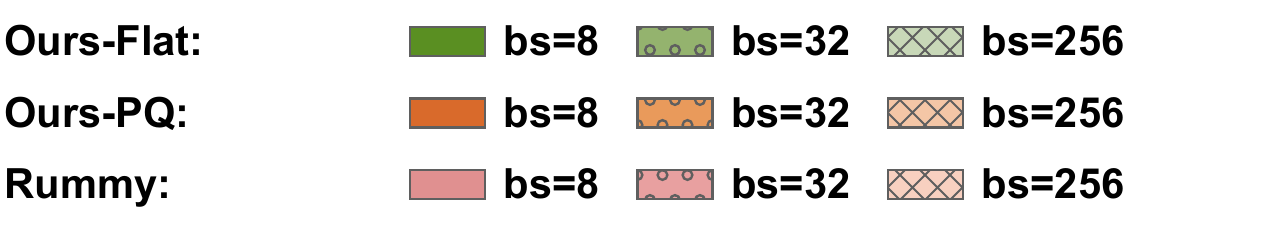}

    \setlength{\tabcolsep}{1pt}
    \renewcommand{\arraystretch}{1.0}
    \begin{tabular}{c c @{\hspace{0.35em}} c @{\hspace{0.35em}} c}
        &
        \hspace*{1.0em}\small\textbf{SIFT-1B} &
        \hspace*{1.0em}\small\textbf{DEEP-1B} &
        \hspace*{1.0em}\small\textbf{SPACEV-1B}
        \\[0.2em]

        \raisebox{-0.4\height}{\rotatebox{90}{\small\textbf{Ours-Flat}}} &
        \raisebox{-0.5\height}{\includegraphics[width=0.1\textwidth]{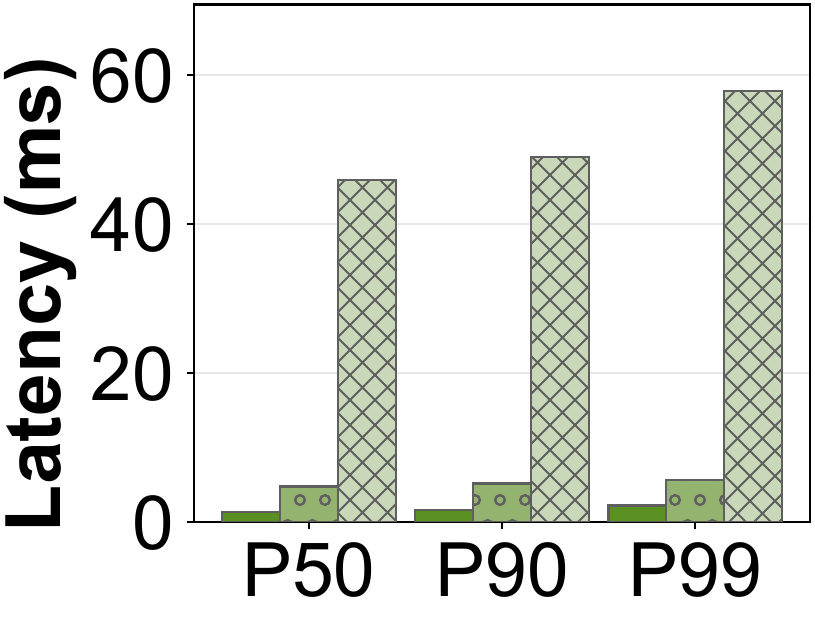}} &
        \raisebox{-0.5\height}{\includegraphics[width=0.1\textwidth]{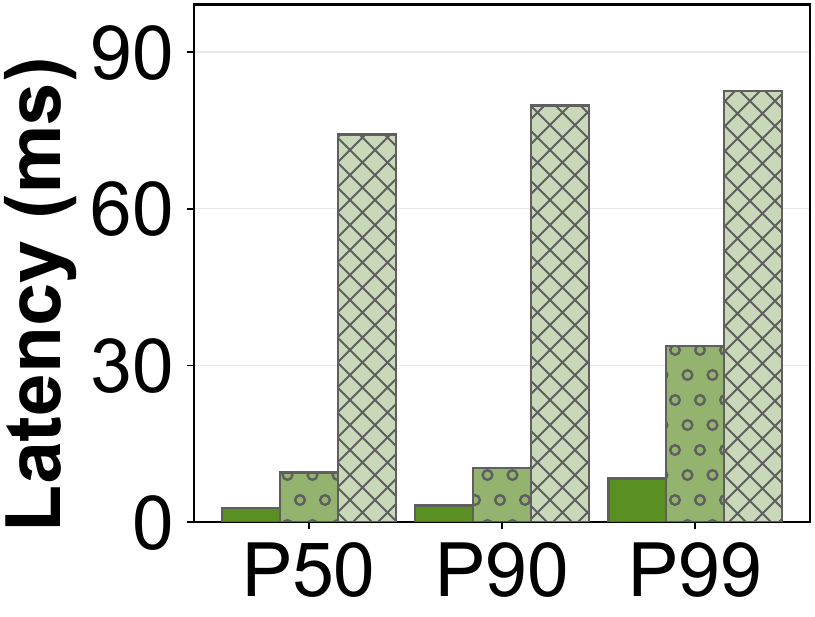}} &
        \raisebox{-0.5\height}{\includegraphics[width=0.1\textwidth]{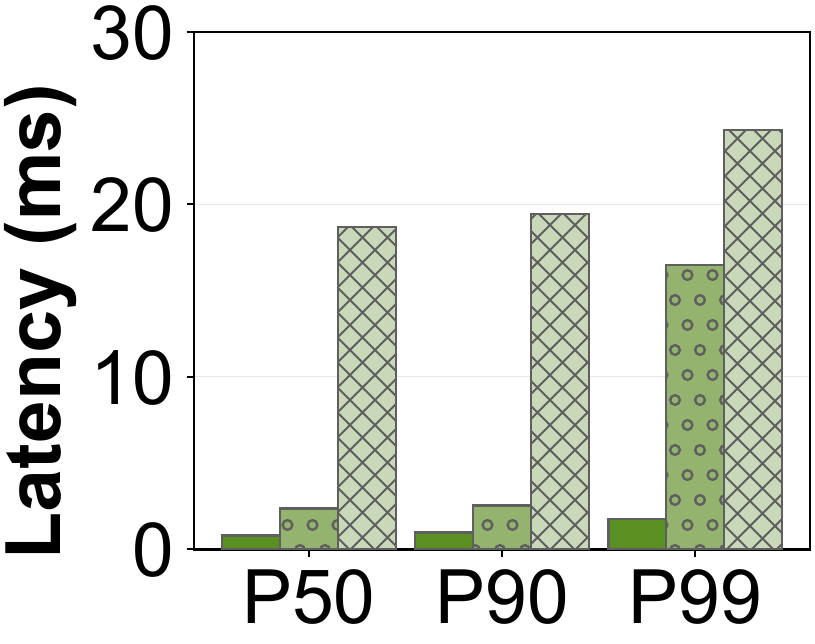}}
        \\[0.0em]

        \raisebox{-0.4\height}{\rotatebox{90}{\small\textbf{Ours-PQ}}} &
        \raisebox{-0.5\height}{\includegraphics[width=0.1\textwidth]{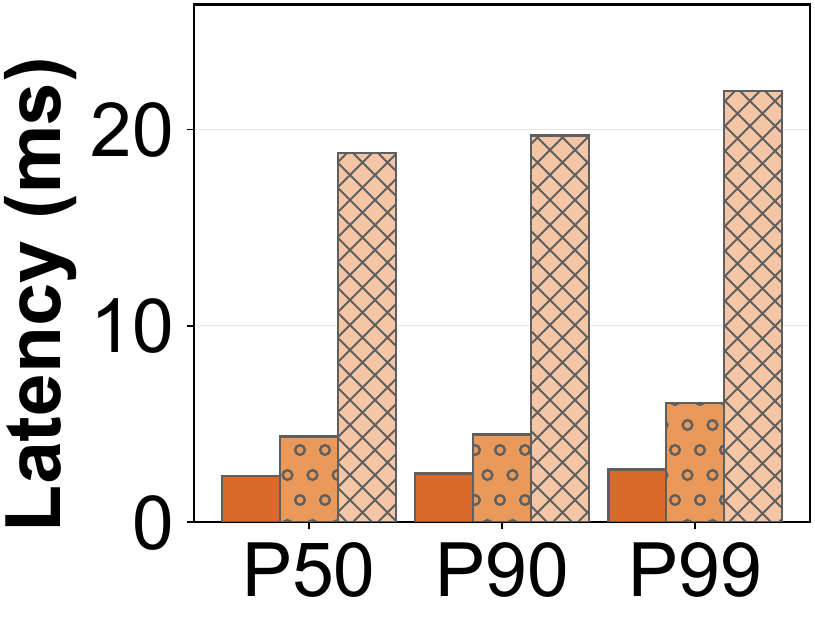}} &
        \raisebox{-0.5\height}{\includegraphics[width=0.1\textwidth]{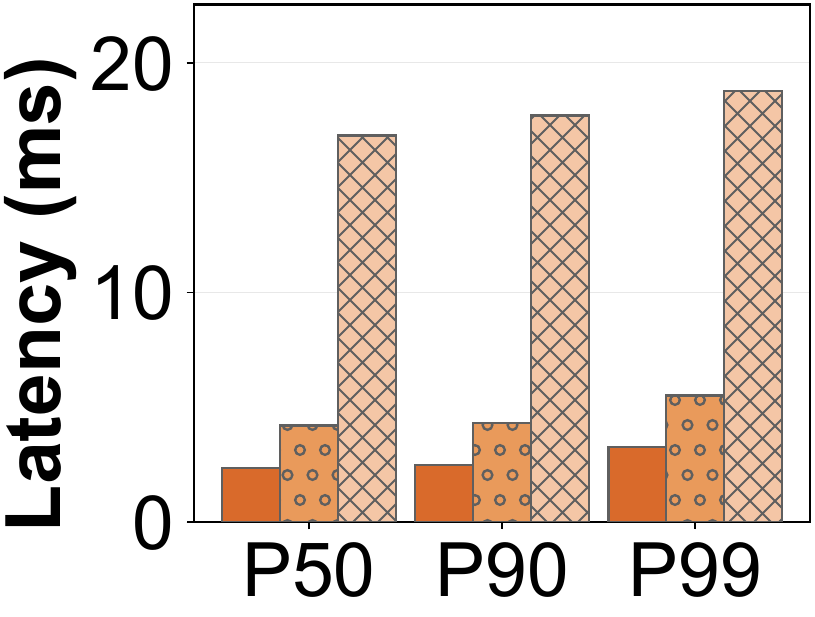}} &
        \raisebox{-0.5\height}{\includegraphics[width=0.1\textwidth]{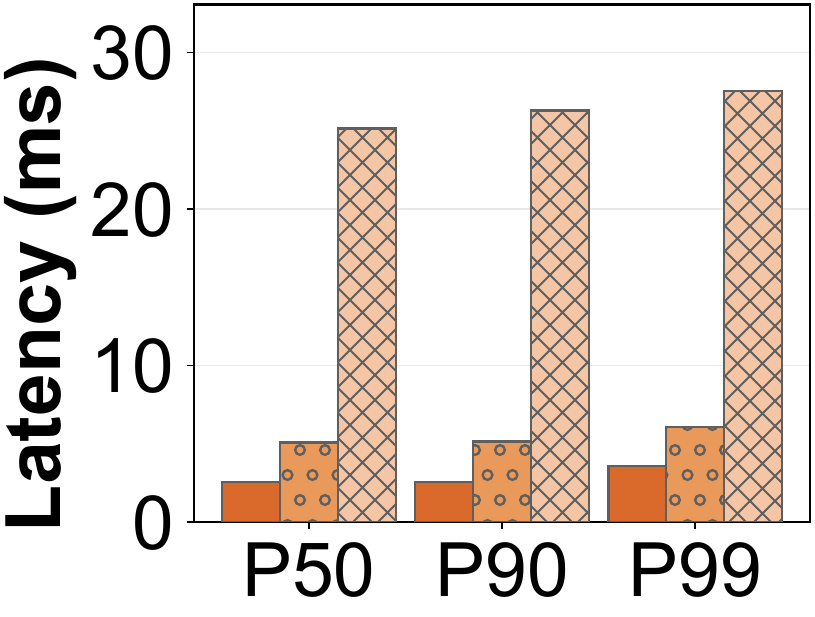}}
        \\[0.0em]

        \raisebox{-0.4\height}{\rotatebox{90}{\small\textbf{Rummy}}} &
        \raisebox{-0.5\height}{\includegraphics[width=0.1\textwidth]{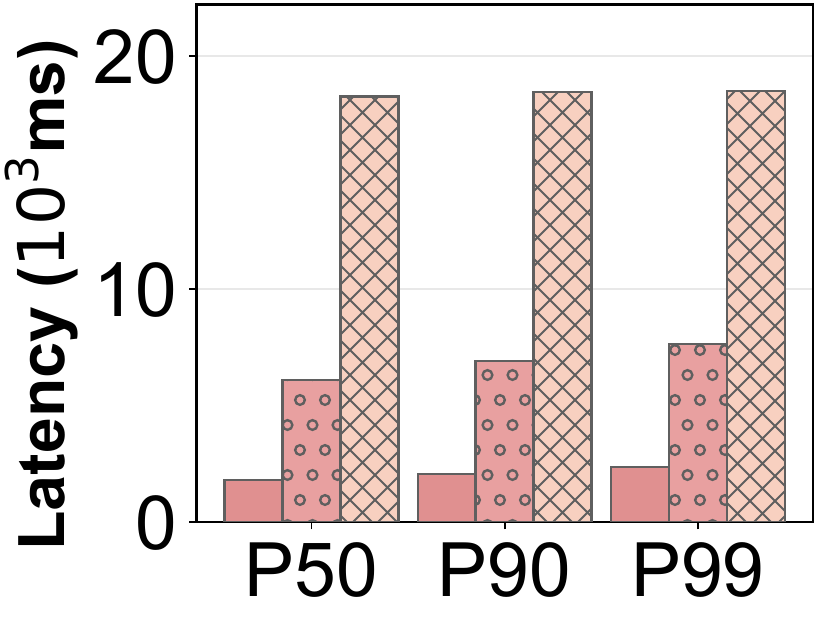}} &
        \raisebox{-0.5\height}{\includegraphics[width=0.1\textwidth]{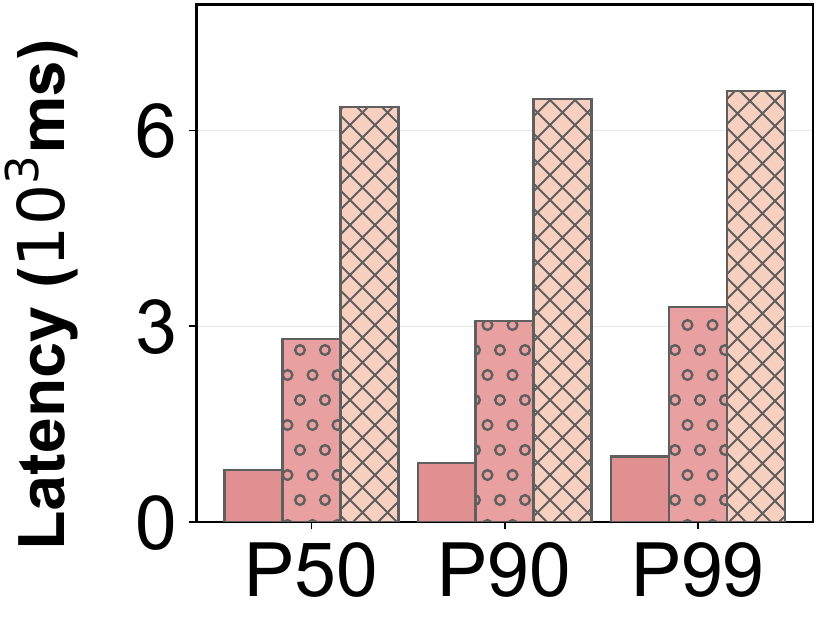}} &
        \raisebox{-0.5\height}{\includegraphics[width=0.1\textwidth]{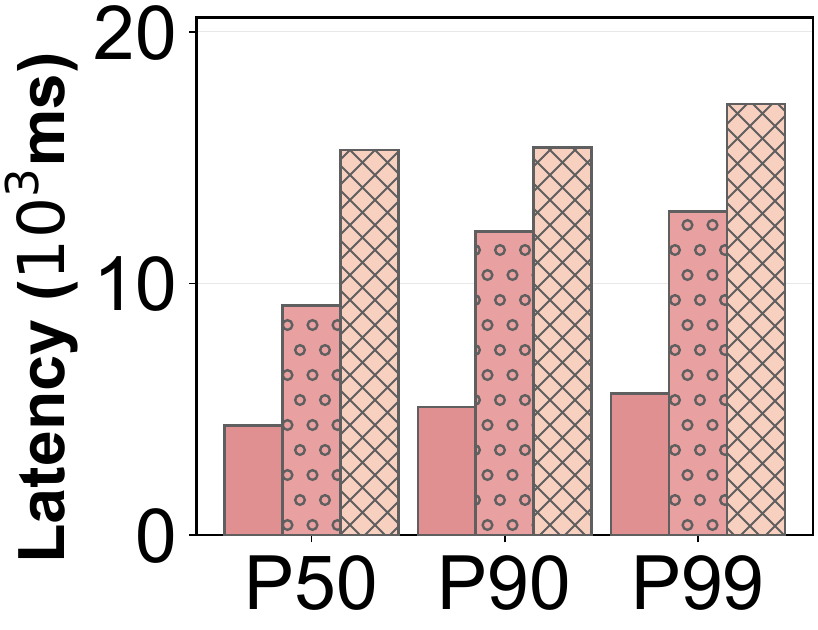}}
    \end{tabular}

    \caption{P50/P90/P99 batch latencies at Recall@10$\approx$0.9.}
    \label{fig:latency_dist_bar}
\end{figure}

Figure~\ref{fig:latency_dist_bar} reports the P50, P90, and P99 batch latencies
of \ourname{} and Rummy. Across all datasets, Rummy reaches the $10^3$-ms scale
while \ourname{} remains in the millisecond range, consistent with the PCIe
cost of moving Rummy's candidate vectors to the GPU. \ourname{} instead avoids
bulk candidate-vector transfers by scanning host-resident vectors in situ and
returning only compact candidate identifiers for subsequent processing.

\begin{figure*}[t]
    \centering
    \includegraphics[width=0.80\textwidth]{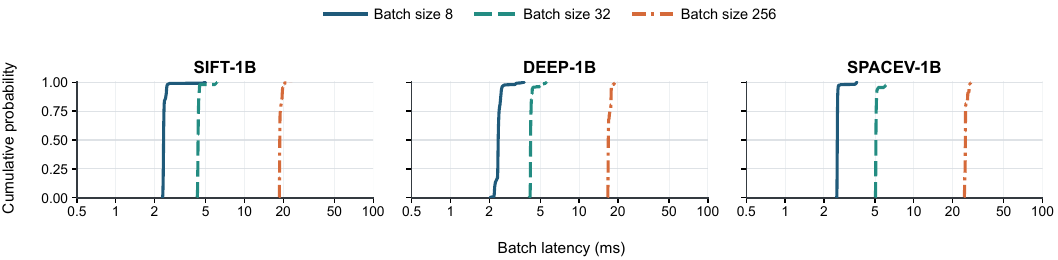}
    \caption{Batch-latency ECDFs of \ourname{}-PQ at
    Recall@10$\approx$0.9.}
    \label{fig:latency_dist_cdf}
    \vspace{0em}

    \includegraphics[width=0.80\textwidth]{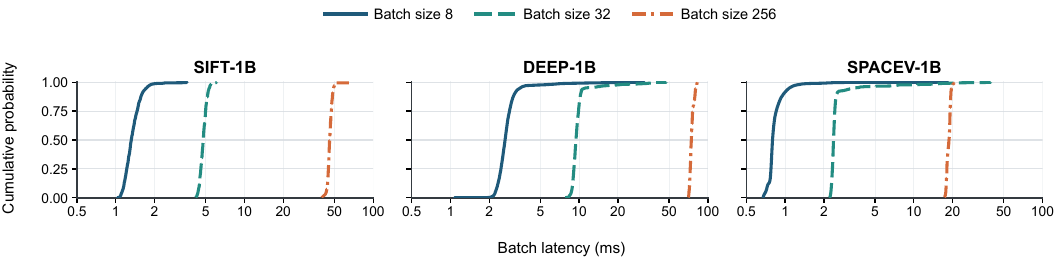}
    \caption{Batch-latency ECDFs of \ourname{}-Flat at
    Recall@10$\approx$0.9.}
    \label{fig:latency_dist_cdf_flat}
\end{figure*}

To examine the distribution beyond these three percentiles,
Figure~\ref{fig:latency_dist_cdf} reports empirical CDFs of the measured batch
latency for \ourname{}-PQ on the three billion-scale datasets. Increasing the
batch size shifts each distribution to the right, as expected because each batch
contains more queries. The PQ distributions remain comparatively compact across
datasets and batch sizes because PQ filtering bounds the amount of subsequent
exact re-ranking. The ECDF figures use a logarithmic latency axis and the
steady-state filtering described above.

Figure~\ref{fig:latency_dist_cdf_flat} reports the corresponding
\ourname{}-Flat ECDFs under the same steady-state filtering protocol. The Flat
distributions are compact on SIFT-1B and at batch size 256, but show visible
upper tails on DEEP-1B and SPACEV-1B at batch size 32, consistent with
query-dependent variation in host-side list scanning.

Together, these results show that both paths avoid Rummy's PCIe-induced
second-scale latency. The PQ path provides tighter latency predictability
through bounded exact re-ranking, whereas the Flat path is more sensitive to
query-dependent variation in the probed IVF lists.

\subsection{Bottleneck Analysis}
\label{subsec:bottleneck}
\begin{figure*}
    \centering
    \includegraphics[width=0.98\textwidth]{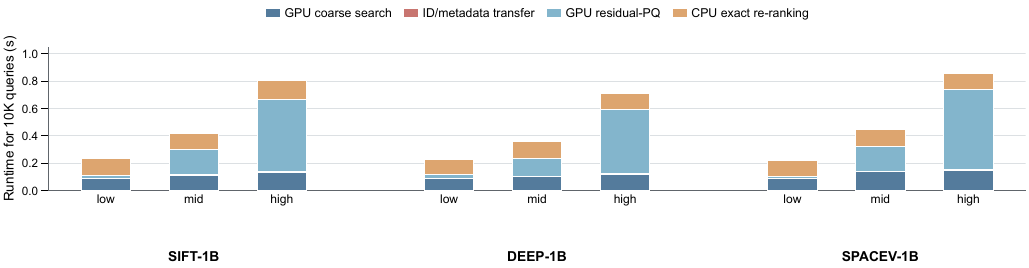}
    \caption{Stage-time breakdown of \ourname{}-PQ.}
    \vspace{-1.1em}
    \label{fig:detailed-breakdown}
\end{figure*}

Figure~\ref{fig:detailed-breakdown} reports the no-pipeline additive breakdown
of \ourname{}-PQ at batch size 2048. Its bars are audited 10K-query runs, and
low/mid/high denote the first, median, and last points of each sweep.
Appendix~\ref{app:stage-breakdown} gives the corresponding Flat and batch-8
breakdowns in Figures~\ref{fig:detailed-breakdown-flat}
and~\ref{fig:detailed-breakdown-bs8}.
Each stack sums the measured query-path stages from the same no-pipeline run.
ID/metadata transfer remains small throughout the PQ sweep. At comparable
recall on SIFT-1B with $\mathrm{bs}=2048$, Rummy at $P=4$ spends about 10~s on
PCIe (Recall@10=0.9147), whereas \ourname{}-PQ at $P=96$ spends only 3.0~ms on
ID/metadata transfer (Recall@10=0.9191). Across all three datasets, residual-PQ
time rises from low to high recall as GPU filtering absorbs the additional
candidate work, while bounded CPU exact re-ranking remains comparatively stable.

\begin{figure}
\centering
\includegraphics[width=\columnwidth]{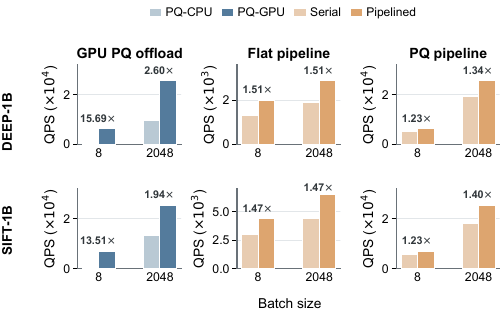}
\caption{Ablation throughput.}
\vspace{-1.0em}
\label{fig:ablation-1b}
\end{figure}

\subsection{Ablations}

\label{subsec:ablation_study}

Figure~\ref{fig:ablation-1b} reports throughput at matched high recall
(Recall@10$\approx$0.95); labels show speedup over each paired baseline. We
ablate two latency-critical choices in \ourname{}: GPU offloading of residual-PQ pruning and pipelining between GPU filtering and CPU verification. PQ-CPU performs the
residual-PQ LUT scan, candidate merging, and exact re-ranking on the CPU,
whereas PQ-GPU moves the PQ scan and candidate merging to the GPU, leaving
only bounded exact re-ranking on the CPU.

GPU offloading removes the dominant bottleneck at small batches. At batch size
8, PQ-GPU is 13.5$\times$ faster than PQ-CPU on SIFT-1B and 15.7$\times$
faster on DEEP-1B. The gain remains substantial even at batch size 2048, where
PQ-GPU still improves throughput by 1.9$\times$ on SIFT-1B and 2.6$\times$ on
DEEP-1B. This shows that CPU-side residual-PQ pruning is a major limiter even
outside the low-batch regime.

Pipelining provides an additional gain by hiding the remaining CPU exact
verification. For \ourname{}-Flat, pipelining delivers a consistent $\sim$1.5$\times$
throughput improvement across batch sizes 8 and 2048 on both SIFT-1B and
DEEP-1B.
For \ourname{}-PQ, it further adds 1.2$\times$ speedup at batch size 8 and
1.3--1.4$\times$ at batch size 2048. Overall, GPU offloading removes most of
the PQ pruning cost from the CPU critical path, while pipelining hides the
bounded CPU verification that remains.

\subsection{Fast Index Construction}
\label{sec:eval:fast-build}

Building a billion-scale IVF index is costly for two reasons: centroid training
repeatedly assigns training vectors, and the completed centroids must then
assign every base vector to an inverted list. We reduce both costs. First, the
fast modes train $N_{\mathrm{list}}=524{,}288$ centroids on a 5M subset rather
than the full base dataset. Second, graph-assisted assignment uses a
CAGRA graph~\cite{ootomo2024cagra} over the current centroids to retrieve a
small candidate set for exact distance evaluation during both training and
full-dataset assignment. Ultra-fast further reduces the Lloyd iterations and
graph-search budget while using larger batches.

\begin{figure}
\centering
\includegraphics[width=0.95\columnwidth]{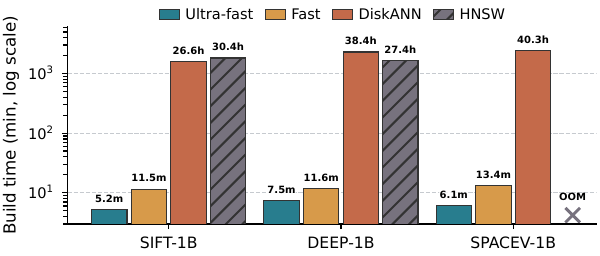}
\caption{Index construction time across billion-scale datasets.}
\vspace{-1.0em}
\label{fig:fast-build-time}
\end{figure}

We measure end-to-end construction wall time on the primary RTX 5090 server
described above, using 64 CPU threads.
Figure~\ref{fig:fast-build-time} compares the two modes with DiskANN and Faiss HNSW
($M_{\mathrm{HNSW}}=16$, $ef_{\mathrm{construction}}=100$). Ultra-fast
finishes SIFT-1B, DEEP-1B, and SPACEV-1B in 5.2, 7.5, and 6.1 minutes;
fast finishes them in 11.5, 11.6, and 13.4 minutes. HNSW takes 30.4 and
27.4 hours on SIFT-1B and DEEP-1B, respectively, and fails with OOM after
31.9 hours at 72.8\% of SPACEV-1B. On the two completed HNSW builds, the
accelerated modes are about $142\times$--$350\times$ faster. Both modes retain
the same final IVF layout and online search path.

\textbf{Construction configurations.}
Fast applies both subset training and graph-assisted assignment. Ultra-fast
uses the same two techniques but reduces the number of Lloyd iterations and
the construction-time CAGRA search budget while increasing the training and
assignment batch sizes. This trades a small amount of routing quality for lower
construction time without changing the serving-time kernels or final index
layout. Table~\ref{tab:fast-build-config-resource} summarizes the configurations.

\begin{table}[t]
\centering
\caption{Accelerated index-construction parameters.}
\label{tab:fast-build-config-resource}
\resizebox{\columnwidth}{!}{%
\begin{tabular}{lcc}
\toprule
Component & Fast & Ultra-fast \\
\midrule
Iterations & 5 & 3 \\
Training batch size & 131,072 & 524,288 \\
Assignment batch size & 65,536 & 1,048,576 \\
Candidate centroids per vector & 8 & 2 \\
CAGRA \texttt{itopk\_size} & 96 & 32 \\
CAGRA graph degree & 32 & 16 \\
CAGRA intermediate graph degree & 64 & 32 \\
\midrule
Final CAGRA graph VRAM & 0.067 GB & 0.034 GB \\
\bottomrule
\end{tabular}}
\end{table}

\textbf{Stage and resource details.}
Table~\ref{tab:fast-build-timing} reports the total end-to-end wall time for
each construction method. The completed SIFT-1B and DEEP-1B
HNSW builds peak at 890 and 982~GB RSS and produce approximately 612 and
492~GB indexes, respectively. On SPACEV-1B, HNSW reaches 1.02B of 1.402B
vectors (72.8\%) before the memory limit terminates the process; it runs for
31.9 hours, peaks at 862~GB RSS, and produces no completed index.
A separate 64-thread DiskANN build with $R=64$ and $L_{\mathrm{build}}=200$
takes 26.6, 38.4, and 40.3 hours on SIFT-1B, DEEP-1B, and SPACEV-1B,
respectively.

\begin{table}[t]
\centering
\caption{Construction wall time for accelerated IVF, DiskANN, and HNSW.}
\label{tab:fast-build-timing}
\resizebox{\columnwidth}{!}{%
\begin{tabular}{lcccc}
\toprule
Dataset & Ultra-fast & Fast & DiskANN & HNSW \\
\midrule
SIFT-1B   & 313 s & 690 s & 26.6 h & 30.4 h \\
DEEP-1B   & 447 s & 696 s & 38.4 h & 27.4 h \\
SPACEV-1B & 364 s & 805 s & 40.3 h & OOM at 72.8\% \\
\bottomrule
\end{tabular}}
\end{table}

CAGRA is built only over the $0.5\mathrm{M}$ IVF centroids and is discarded
after construction; no billion-node graph is retained in the serving index.

\section{Related Work}
\label{sec:related}

ANN search has been studied from both algorithmic~\cite{arya1993approximate,datar2004locality} and systems perspectives~\cite{wang2021milvus,chen2021spann}. 
Graph-based indexes such as HNSW~\cite{malkov2018efficient}, NSG~\cite{fu2017fast}, DiskANN~\cite{jayaram2019diskann}, and CAGRA~\cite{ootomo2024cagra} achieve strong recall--latency trade-offs by navigating proximity graphs. 
These methods are widely used for high-recall and latency-sensitive ANN search. 
However, graph indexes often require substantial index memory and relatively expensive construction or rebuild procedures, especially at billion scale~\cite{singh2021freshdiskann,manohar2024parlayann}. 
\ourname{} takes a different direction: rather than optimizing graph traversal, it revisits IVF and studies how a simple partition-based index can be mapped efficiently onto a heterogeneous CPU--GPU memory hierarchy. 
This design targets high batch throughput and efficient rebuilds while retaining the regular structure of IVF.

IVF and PQ are classical building blocks for large-scale vector search~\cite{johnson2019billion,baranchuk2018revisiting,jegou2010product,babenko2014inverted}.
Systems such as Faiss~\cite{faiss2024} and recent GPU-accelerated libraries~\cite{johnson2019billion,faiss_cuvs2025} demonstrate the effectiveness of regular, batched vector-search operators. 
These systems are highly efficient when the working set fits in GPU memory or when compressed representations are sufficient for the target workload~\cite{ge2013optimized,kalantidis2014locally,babenko2014additive,zhang2014composite,guo2020accelerating,gao2024rabitq}. 
In contrast, \ourname{} targets host-resident billion-scale datasets by placing
compact routing and pruning on the GPU while performing exact in-situ
verification on the CPU, achieving strong recall--QPS trade-offs without
requiring the full base vectors to fit in GPU memory.

Recent systems explore ANN search beyond GPU memory by coordinating CPU memory,
GPU compute, and host--device transfers~\cite{zhang2024fast,tian2025towards,peng2026svfusion}. 
Auncel scales query processing across distributed workers~\cite{zhang2023fast}, while Rummy and GPU-accelerated database systems transfer candidate data to the GPU and overlap data movement with computation~\cite{zhang2024fast,li2025scaling}.
FusionANNS targets an SSD-resident graph index, combining CPU graph navigation,
GPU compressed-vector filtering, and SSD retrieval of raw vectors; its
optimizations reduce page I/O and read amplification~\cite{tian2025towards}.
\ourname{} instead uses dense IVF centroid routing on the GPU and scans
list-major raw vectors in host DRAM, limiting PCIe traffic to compact metadata
and shifting the critical resource from SSD I/O to DRAM bandwidth~\cite{williams2009roofline}.

\section{Conclusion}
\label{sec:conclusion}

\ourname{} demonstrates that billion-scale vector search can be made fast, practical, and cost-efficient without storing all base vectors on the GPU or relying on heavy graph indexes. By assigning IVF’s coarse search to the GPU, performing fine search in situ over host-resident vectors, and incorporating PQ acceleration, \ourname{} avoids costly PCIe transfers, simultaneously preserves simplicity, update-friendliness, and low memory overhead, and achieves strong recall, high throughput, and ultra-fast index construction on commodity hardware, offering a competitive and scalable alternative to existing GPU-heavy or graph-based ANNS systems.
{
\balance
\bibliographystyle{ACM-Reference-Format}
\bibliography{reference}
}

\clearpage
\appendix

\section{Update Handling}
\label{sec:update}

\ourname{} supports lightweight updates through a snapshot--delta design that
keeps the query-visible IVF layout immutable. The published state consists of a
main snapshot, a list-local delta segment, and a tombstone bitmap. Each query
batch pins one consistent version of these three components, avoiding graph
rewiring or in-place list mutation on the online path.

\textbf{Insertions.}
An inserted vector is assigned with the snapshot's centroid table and appended
to the corresponding list-local buffer. These buffers are periodically
published in cluster-major order, so a probe scans one main range followed by
one delta range without per-vector indirection. For \ourname{}-PQ, the delta
stores full vectors for exact verification and residual-PQ codes produced by the
current codebooks~\cite{jegou2010product}; main and delta candidates therefore
share the same GPU pruning and CPU exact re-ranking path.

\textbf{Tombstone-based deletions.}
A deletion records the vector identifier in the tombstone bitmap rather than
rewriting its IVF list. Search filters tombstoned candidates and performs a
final check during exact verification, ensuring that deleted vectors are not
returned.

\textbf{Background compaction and rebuild.}
When the delta or tombstones become large, a background rebuild folds inserts
into a fresh snapshot, removes deleted vectors, and optionally refreshes the PQ
assets. An atomic swap publishes the validated snapshot; pinned queries finish
on the old version, while later queries use the compacted layout.

\subsection{Update Overhead}

We evaluate balanced insert/delete workloads of 10M, 50M, and 100M vectors,
corresponding to 1\%, 5\%, and 10\% of SIFT-1B. Table~\ref{tab:update-overlay}
reports the resulting changes in recall, throughput, and batch-latency
percentiles.

Search quality remains stable as the update set grows. The largest absolute
Recall@10 changes are 0.001254 for \ourname{}-Flat and 0.001940 for
\ourname{}-PQ, both at the 100M setting. This confirms that searching the main
snapshot and published delta together preserves the result quality of the
corresponding no-update index.

The query-time cost depends on the search path and batch size. \ourname{}-Flat
adds one host-memory delta range per probed list, giving 3.46\%--7.10\%
throughput overhead at batch size 8 and only 0.57\%--1.14\% at batch size 2048.
For \ourname{}-PQ, the published residual codes add work to GPU pruning before
the bounded exact re-ranking stage. Its overhead is 21.62\%--22.63\% at batch
size 8 and falls to 4.72\%--7.42\% at batch size 2048 as fixed scan and merge
costs are better amortized. The P50 and P99 changes follow the same pattern:
small-batch PQ is the most sensitive setting, whereas both paths are more stable
at the larger batch size.

Overall, the snapshot--delta design keeps recall changes small without online
structural mutation. The Flat path has the lowest query-time update overhead,
while the PQ path retains a unified pruning and verification pipeline for main
and delta data at the cost of additional residual-code scanning.

\begin{table}[!b]
\centering
\small
\caption{
Query-time overhead of lightweight updates on SIFT-1B. Throughput overhead is computed
as $1-\mathrm{QPS}_{update}/\mathrm{QPS}_{base}$, where
the base is the no-update baseline.
}
\label{tab:update-overlay}
\resizebox{\columnwidth}{!}{
\begin{tabular}{llrrrr}
\toprule
\textbf{\ourname{} mode} & \textbf{BS} & \textbf{Insert/Delete}
& \textbf{QPS Overhead} & $\mathbf{|\Delta R@10|}$
& \textbf{$\Delta$P50 / $\Delta$P99} \\
\midrule
Flat & 8    & 10M/10M   & 3.46\% & 0.000150 & 4.28\% / -0.77\% \\
Flat & 8    & 50M/50M   & 7.10\% & 0.000669 & 6.05\% / 3.34\% \\
Flat & 8    & 100M/100M & 6.17\% & 0.001254 & 5.04\% / 6.18\% \\
Flat & 2048 & 10M/10M   & 0.57\% & 0.000140 & 0.64\% / 0.70\% \\
Flat & 2048 & 50M/50M   & 1.14\% & 0.000660 & 0.83\% / 1.47\% \\
Flat & 2048 & 100M/100M & 0.84\% & 0.001243 & 0.88\% / 0.82\% \\
\midrule
PQ    & 8    & 10M/10M   & 21.73\% & 0.000260 & 26.95\% / 21.15\% \\
PQ    & 8    & 50M/50M   & 22.63\% & 0.000880 & 27.13\% / 19.90\% \\
PQ    & 8    & 100M/100M & 21.62\% & 0.001930 & 26.67\% / 8.85\% \\
PQ    & 2048 & 10M/10M   & 7.42\% & 0.000260 & 6.90\% / 7.43\% \\
PQ    & 2048 & 50M/50M   & 5.94\% & 0.000880 & 5.64\% / 5.04\% \\
PQ    & 2048 & 100M/100M & 4.72\% & 0.001940 & 4.14\% / 4.01\% \\
\bottomrule
\end{tabular}}
\end{table}

\section{Evaluation Protocol and Baseline Details}
\label{app:evaluation-details}

\textbf{Host execution and memory use.}
The large DRAM capacity holds host-resident vectors and supports
memory-intensive offline DiskANN/HNSW construction. We do not maintain extra
serving-time copies of the base-vector payload or query-dependent intermediate
results. We leave NUMA placement and thread scheduling to the operating system
and OpenMP runtime. Explicit NUMA binding or manually partitioning IVF lists by
socket does not consistently improve performance in our measurements because
the query-dependent exact fine-search workload benefits from dynamic load
balancing across CPU workers.

\textbf{DiskANN.}
We use an on-disk Vamana index and a patched batched search runner. SIFT-1B uses
$R=64$ and $L_{\mathrm{build}}=200$; DEEP-1B uses the same configuration;
SPACEV-1B uses $R=24$ and $L_{\mathrm{build}}=64$. Search runs in SSD-based
serving mode with beamwidth 2 and 64 threads. The SIFT-1B and DEEP-1B runs do
not cache graph nodes.

\textbf{PipeANN.}
We evaluate PipeANN~\cite{guo2025pipeann} on SIFT-1B and SPACEV-1B with 56 CPU threads and sweep its
search budget $L$ at batch sizes 8, 32, 256, and 2048. Each operating point is
the mean of five runs, and Figure~\ref{fig:graph_compare_1b} reports the
non-dominated Recall@10--QPS frontier.

\textbf{HNSW.}
We use 64 CPU threads and the same batch sizes as \ourname{}. On the 100M
datasets, $M_{\mathrm{HNSW}}=20$ and $ef_{\mathrm{construction}}=100$. At
billion scale, we use the largest configurations that complete without OOM:
$M_{\mathrm{HNSW}}=16$ and $ef_{\mathrm{construction}}=100$ on SIFT-1B and
DEEP-1B. On SPACEV-1B, larger settings exceed available memory or fail to
complete, so we report $M_{\mathrm{HNSW}}=10$ and
$ef_{\mathrm{construction}}=32$ for the search comparison. The controlled
construction study in Section~\ref{sec:eval:fast-build} instead fixes
$M_{\mathrm{HNSW}}=16$ and $ef_{\mathrm{construction}}=100$ across datasets.

\section{Fast Index Building Details}
\label{app:fast-index-building}

For billion-scale ANNS systems, index build time directly affects deployment,
parameter tuning, and snapshot refresh. Standard IVF construction spends most
of this work in two places: iterative centroid training and the final assignment
of every base vector. Our accelerated modes target these costs separately.

\textbf{Subset-based centroid training.}
Instead of clustering the full base set, the fast and ultra-fast modes
train $N_{\mathrm{list}}=524{,}288$ IVF centroids on a 5M sample.
This reduces the number of vector assignments performed across Lloyd iterations
while retaining enough samples to learn the large centroid table.

\textbf{Graph-assisted centroid assignment.}
We build a CAGRA graph~\cite{ootomo2024cagra} over the current centroid table.
For each sampled training vector, the graph retrieves a small candidate-centroid
set, followed by exact GPU distance evaluation within that set. After training,
the same graph-assisted procedure assigns the full base dataset in batches.
Thus, the graph reduces centroid-search work in both construction stages and is
not part of the serving-time index.

\textbf{Construction modes.}
The full mode is the full-data, non-graph-assisted methodological reference.
Fast mode applies both subset training and graph-assisted assignment.
Ultra-fast uses the same two techniques but reduces
the number of Lloyd iterations and
uses a smaller construction-time CAGRA search budget~\cite{ootomo2024cagra}, while increasing the
training and assignment batch sizes. This intentionally trades a small amount
of routing quality for lower construction time, without changing the online
search kernels or the final index layout. Table~\ref{tab:app-fast-build-config-resource}
summarizes the two accelerated configurations.

\textbf{Stage and end-to-end timings.}
All InSituANN and HNSW construction results use end-to-end wall time on the
primary RTX 5090 server described in Section~\ref{sec:eval}, using 64 CPU
threads. The DiskANN reference uses a separate 64-thread build with $R=64$
and $L_{\mathrm{build}}=200$. Table~\ref{tab:app-fast-build-timing} reports
the resulting wall times. Across the three datasets,
ultra-fast completes in 5.2--7.5 minutes and fast in 11.5--13.4 minutes.

\begin{table}[H]
\centering
\caption{Construction wall time for accelerated IVF, DiskANN, and HNSW.}
\label{tab:app-fast-build-timing}
\resizebox{\columnwidth}{!}{%
\begin{tabular}{lcccc}
\toprule
Dataset & Ultra-fast & Fast & DiskANN & HNSW \\
\midrule
SIFT-1B   & 313 s & 690 s & 26.6 h & 30.4 h \\
DEEP-1B   & 447 s & 696 s & 38.4 h & 27.4 h \\
SPACEV-1B & 364 s & 805 s & 40.3 h & OOM at 72.8\% \\
\bottomrule
\end{tabular}}
\end{table}

The HNSW baseline uses Faiss \texttt{IndexHNSWFlat} with
$M_{\mathrm{HNSW}}=16$, $ef_{\mathrm{construction}}=100$, and 64 threads.
The completed SIFT-1B and DEEP-1B builds peak at 890 and 982~GB RSS and produce
approximately 612 and 492~GB indexes, respectively. On SPACEV-1B, HNSW reaches
1.02B of 1.402B vectors (72.8\%) before the memory limit terminates the process;
it runs for 31.9 hours, peaks at 862~GB RSS, and produces no completed index.

\textbf{CAGRA overhead.}
CAGRA is built only over the $0.5\mathrm{M}$ IVF centroids and is discarded
after construction; no billion-node graph is retained in the serving index.

\begin{table}[H]
\centering
\caption{Construction parameters and GPU-memory footprints for accelerated SIFT-1B index construction.}
\label{tab:app-fast-build-config-resource}
\resizebox{\columnwidth}{!}{
\begin{tabular}{lcc}
\toprule
Component & Fast mode & Ultra-fast mode \\
\midrule
Iterations & 5 & 3 \\
Training batch size & 131,072 & 524,288 \\
Assignment batch size & 65,536 & 1,048,576 \\
Candidate centroids per vector (\texttt{topk\_centroids}) & 8 & 2 \\
CAGRA \texttt{itopk\_size} & 96 & 32 \\
CAGRA graph degree & 32 & 16 \\
CAGRA intermediate graph degree & 64 & 32 \\
\midrule
Final CAGRA graph VRAM & 0.067 GB & 0.034 GB \\
\bottomrule
\end{tabular}}
\end{table}

\section{Fine-Search Locality and GPU Caching}
\label{app:access-locality}

The CPU in-situ placement avoids moving original vectors during fine search.
A natural alternative is to cache frequently accessed IVF lists in GPU memory
and stream misses over PCIe. We first check whether accesses are concentrated
enough to justify caching, then test the alternative end to end.

\begin{figure*}[t]
    \centering
    \includegraphics[width=0.68\textwidth]{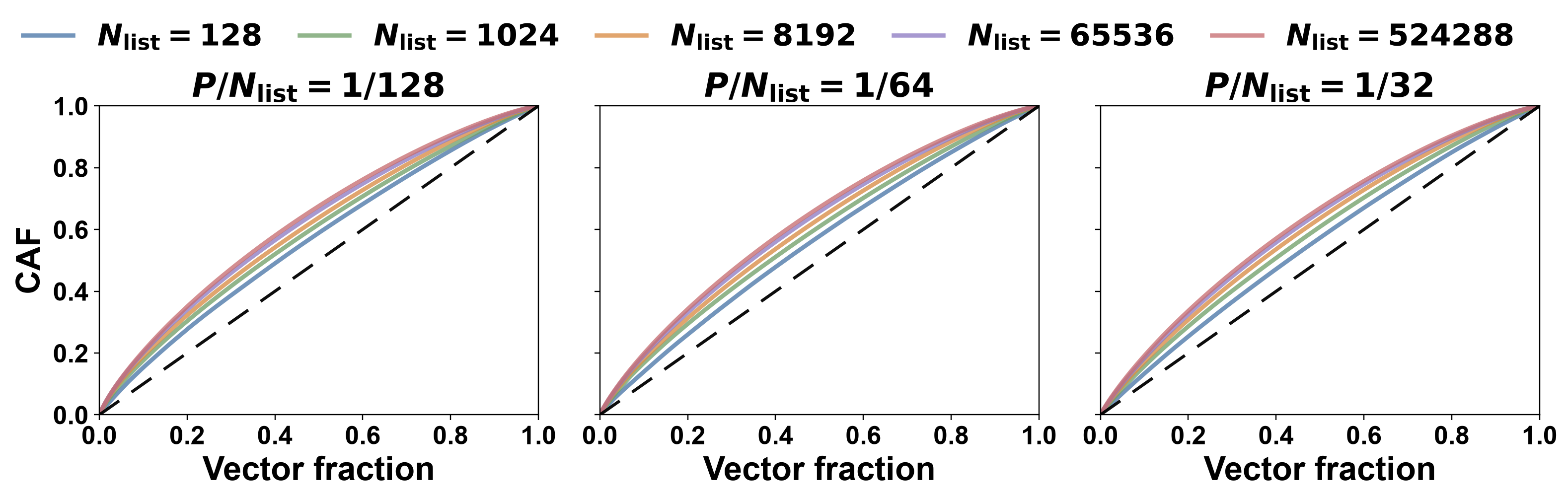}
    \caption{Cumulative fine-search vector accesses on SIFT-100M.}
    \label{fig:access-concentration}
\end{figure*}

\textbf{Access concentration.}
Figure~\ref{fig:access-concentration} shows that access frequencies are only
moderately skewed across list counts and probe ratios. The curves stay close to
the uniform-access diagonal, so no small hot set captures most fine-search
accesses on SIFT-100M.

\textbf{Comparison paths.}
We evaluate three backends under the same IVF routing and candidate set:

\noindent\hspace*{2em}CPU in-situ: keeps the selected IVF lists in host memory and computes
exact distances on the CPU.\par
\noindent\hspace*{2em}GPU uncached: transfers the selected lists to the GPU and performs
exact fine search there.\par
\noindent\hspace*{2em}GPU cache: adds a static, list-granular GPU cache to GPU uncached,
with uncached lists following the same streaming path.\par

Cache contents are selected on a calibration query set and evaluated on a
disjoint held-out set across multiple cache capacities, recall targets, and
batch sizes.

\begin{figure*}[t]
    \centering
    \includegraphics[width=0.72\textwidth]{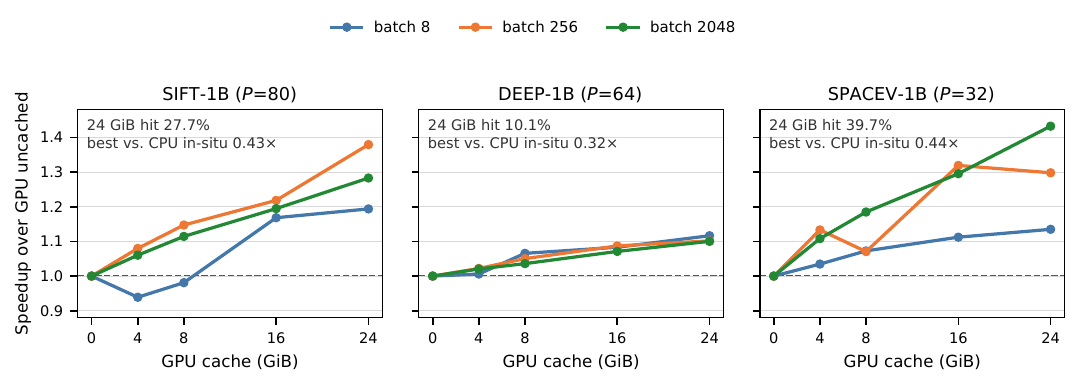}
    \caption{GPU-cache capacity sensitivity for GPU fine search.}
    \label{fig:gpu-cache-capacity}
\end{figure*}

\textbf{Results and implication.}
At Recall@10 $\approx 0.90$ and batch size 2048, a 24~GiB GPU cache improves
GPU uncached throughput by $1.10\times$--$1.43\times$ across the three
datasets. The gains track byte hit rates of 10.1\%--39.7\%; DEEP-1B benefits
least because its larger vectors reduce cache coverage at a fixed capacity
(Figure~\ref{fig:gpu-cache-capacity}).

\section{Sensitivity to the Number of IVF Lists}
\label{app:nlist-sensitivity}
Figure~\ref{fig:nlist-sensitivity} reports the throughput--recall trade-off under different $N_{\mathrm{list}}$ values on SIFT-1B.
The results show that $N_{\mathrm{list}}$ has a non-monotonic effect on performance.
Too few lists make the IVF partitioning too coarse, so each probed cluster contains more irrelevant vectors, increasing the fine-search cost. Too many lists make the IVF partitioning overly fine-grained, so the GPU must search over more centroids, increasing coarse-search overhead.
$N_{\mathrm{list}}=0.5\mathrm{M}$ and $1\mathrm{M}$ achieve the best balance.
They provide high throughput across most recall targets, while larger settings such as $2\mathrm{M}$ and $4\mathrm{M}$ suffer from excessive coarse-search cost.
This suggests that \ourname{} benefits from moderately fine IVF partitioning, but overly large $N_{\mathrm{list}}$ is not worthwhile.

\setcounter{figure}{14}
\begin{figure*}[!b]
    \centering
    \includegraphics[width=0.76\textwidth]{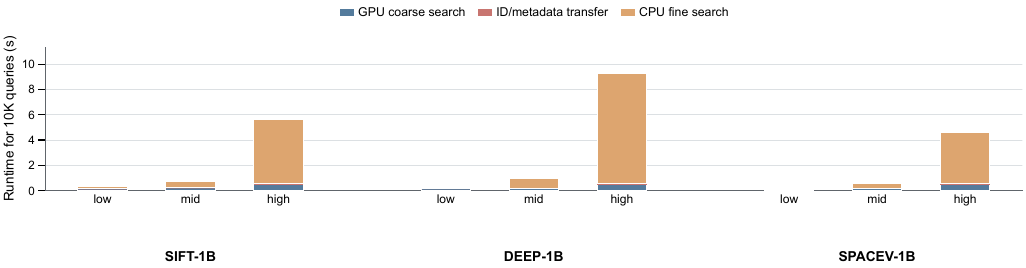}
    \caption{No-pipeline additive stage times for \ourname{}-Flat at batch
    size 2048.}
    \label{fig:detailed-breakdown-flat}
\end{figure*}
\setcounter{figure}{13}

\begin{figure}
\centering
\includegraphics[width=0.86\columnwidth]{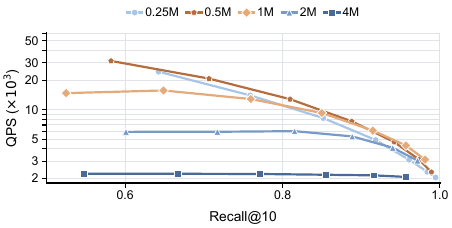}
\caption{Sensitivity to the number of IVF lists on SIFT-1B.}
\label{fig:nlist-sensitivity}
\end{figure}

\section{100M-Scale Results}
\label{app:results-100m}

We include PilotANN on the 100M datasets with
$n_{\mathrm{neighbors}}=32$, $ef_{\mathrm{construction}}=40$, an
$ef_{\mathrm{search}}$ sweep, sample ratio $0.25$, and projected dimensionality
$64$. Under this setting, its GPU-resident staged index requires at least
13.8~GB at 100M vectors, but scales to 138.0~GB for 1B SIFT/DEEP and
193.5~GB for SPACEV-1B, before temporary buffers and allocator overhead. This far exceeds our 32~GB single-GPU memory budget, so PilotANN is reported only on the
100M datasets.

Figure~\ref{fig:graph_compare_100m} shows that the 1B conclusions also hold at 100M scale, where PilotANN can be included as an additional stronger GPU graph baseline.
\ourname{} again gives a broad recall--throughput frontier: \ourname{}-PQ improves throughput over Rummy by about $63.0\times$--$489.5\times$ for $\mathrm{bs}=8,32$ and $6.8\times$--$207.1\times$ for $\mathrm{bs}=256,2048$.
The graph-baseline trend is also preserved: compared with PilotANN, \ourname{}-PQ achieves higher throughput in most settings, with speedups up to $6.3\times$, especially in the high-recall, small-batch regime; against DiskANN, \ourname{}-PQ improves throughput by about $2.2\times$--$33.5\times$ overall.
For HNSW, \ourname{} keeps the stronger high-recall frontier, e.g., about $1.2\times$--$11.9\times$ faster than HNSW on SIFT-100M above Recall@10 $0.9$ and up to $31.9\times$ on DEEP-100M.

\section{Hardware Portability}
\label{app:hardware-portability}
To evaluate hardware portability, we additionally run \ourname{} on a single-H100 node with 475~GB of host DRAM.
This setting is more constrained on the host side than our main evaluation server, stressing the host-resident fine-search design.
Figure~\ref{fig:H100-main} shows that \ourname{} preserves similar
recall--throughput trends on the H100 platform. For \ourname{}-Flat, throughput is primarily
limited by host-side exact distance evaluation rather than GPU compute: at
Recall@10 $\approx 0.91$, it achieves 5.2K, 7.1K, 8.2K, and 8.6K QPS for batch
sizes 8, 32, 256, and 2048, respectively. This is consistent with the RTX 5090
results, showing that upgrading the GPU alone does not substantially accelerate when the bottleneck is host-memory scanning. \ourname{}-PQ benefits more from the H100 at larger batches, since
candidate pruning is offloaded to the GPU and CPU-side scan pressure is
reduced. At Recall@10 $\approx 0.90$, \ourname{}-PQ reaches 21.4K QPS for batch size
256 and 25.9K QPS for batch size 2048, exceeding the corresponding RTX 5090
results. Figure~\ref{fig:h100-exact-latency-percentiles} further shows stable
batch latency for \ourname{}-Flat at Recall@10 $\approx 0.9$, with P99 latencies of
2.02, 6.54, 38.18, and 268.24~ms from batch size 8 to 2048.
These results demonstrate that \ourname{} also achieves strong performance on a single H100 even under constrained host-memory capacity.

\section{Stage Breakdown}
\label{app:stage-breakdown}

\textbf{Flat path at batch size 2048.}
Figure~\ref{fig:detailed-breakdown-flat} reports no-pipeline additive stage
times at representative low, mid, and high operating points, selected as the
first, median, and last points of each sweep.

The Flat breakdown uses the same no-pipeline timing and representative-point
selection as the PQ breakdown. ID/metadata transfer remains a narrow segment,
whereas CPU fine search rises sharply with recall and dominates the high-recall
stack. GPU coarse search grows more slowly. The limiting work is therefore
local host scanning rather than cross-device metadata movement.

The two modes expose different bottleneck structures. In \ourname{}-Flat,
increasing recall mainly amplifies CPU fine search over host-resident full
vectors. In \ourname{}-PQ, the added high-recall work is split between GPU
residual-PQ filtering and bounded CPU exact re-ranking. This makes the PQ path
more controllable at high throughput, while the Flat path remains valuable when
avoiding PQ pruning is necessary for the highest recall.

\textbf{Small batches.}
Figure~\ref{fig:detailed-breakdown-bs8} complements the large-batch PQ and Flat
breakdowns in Figures~\ref{fig:detailed-breakdown}
and~\ref{fig:detailed-breakdown-flat}. It uses the same source sweeps, stage
definitions, normalization, and operating-point selection; only the batch-size
setting differs. The left labels report the measured Recall@10 at each
operating point.

\setcounter{figure}{15}
\begin{figure*}[p]
\centering
\includegraphics[
    width=0.58\textwidth,
    trim=0 2 0 2,
    clip
]{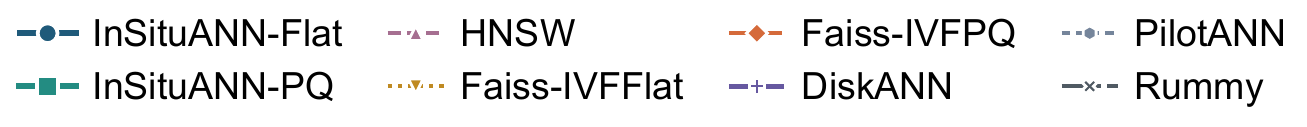}

\resizebox{0.70\textwidth}{!}{%
\setlength{\tabcolsep}{2pt}%
\renewcommand{\arraystretch}{1.0}%
\begin{tabular}{ccccc}
    &
    \hspace*{1.8em}\textbf{batchsize=8} &
    \hspace*{1.8em}\textbf{batchsize=32} &
    \hspace*{1.8em}\textbf{batchsize=256} &
    \hspace*{1.8em}\textbf{batchsize=2048} \\[0.0em]
    \raisebox{-0.4\height}{\rotatebox{90}{\textbf{SIFT-100M}}} &
    \raisebox{-0.5\height}{\includegraphics[width=0.228\textwidth]{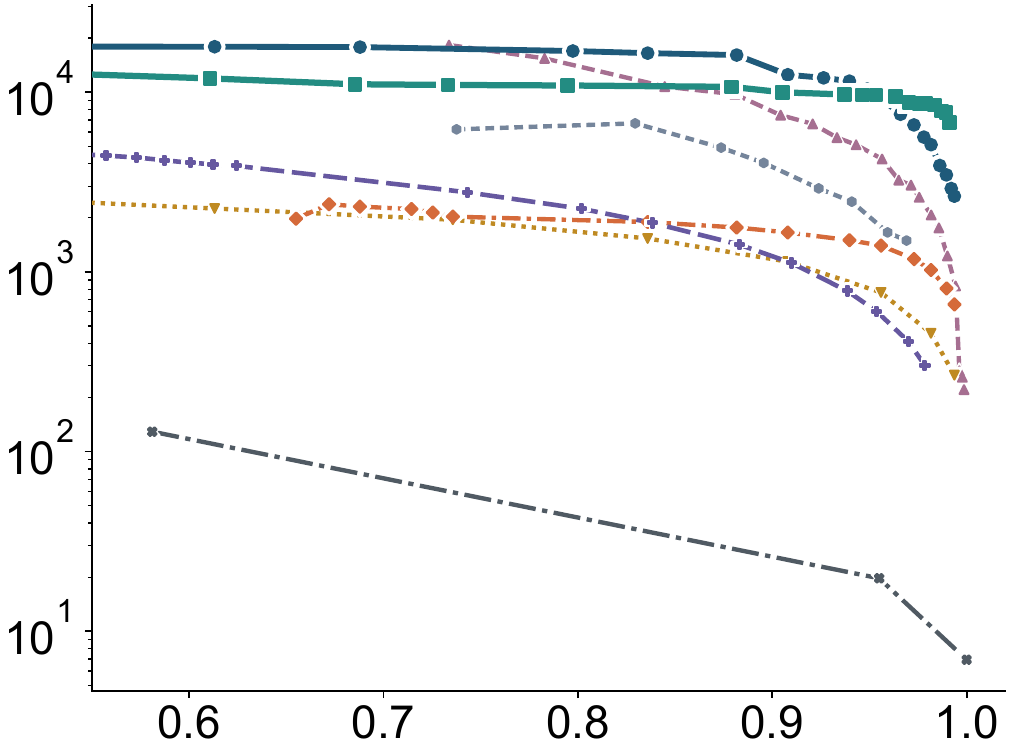}} &
    \raisebox{-0.5\height}{\includegraphics[width=0.228\textwidth]{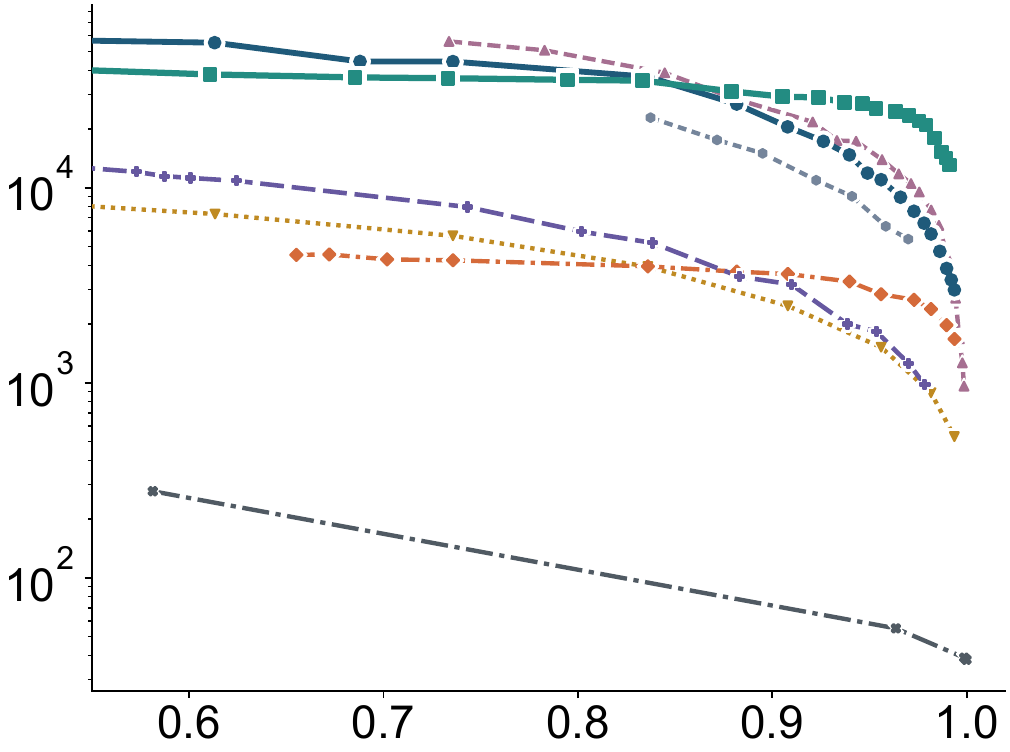}} &
    \raisebox{-0.5\height}{\includegraphics[width=0.228\textwidth]{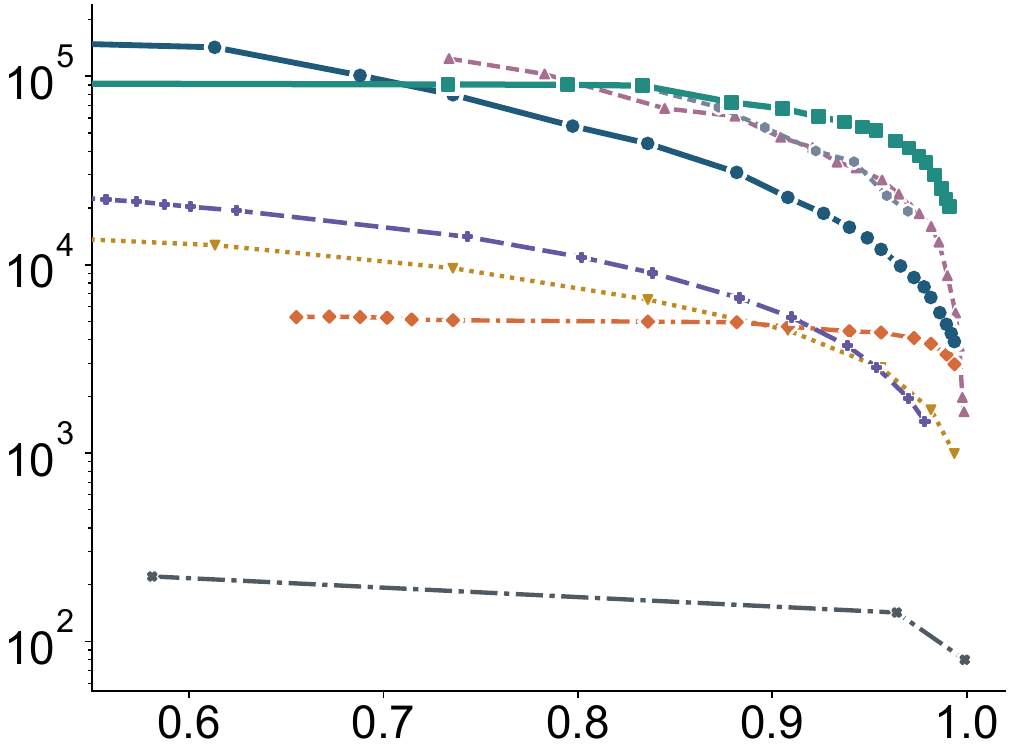}} &
    \raisebox{-0.5\height}{\includegraphics[width=0.228\textwidth]{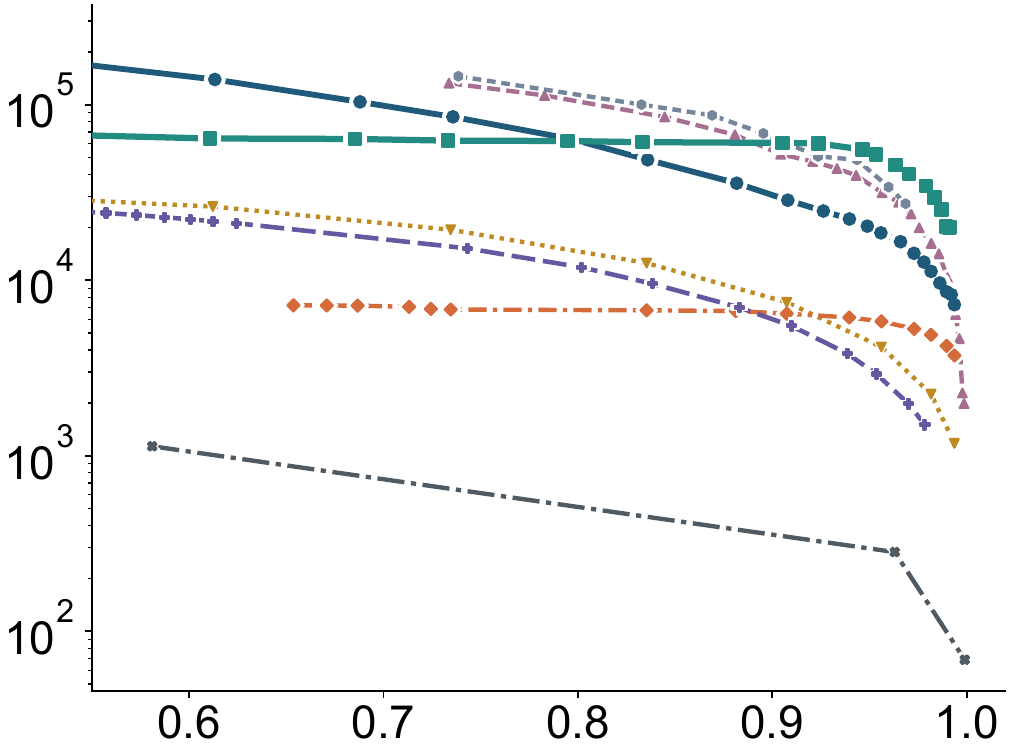}} \\[0.5em]
    \raisebox{-0.4\height}{\rotatebox{90}{\textbf{DEEP-100M}}} &
    \raisebox{-0.5\height}{\includegraphics[width=0.228\textwidth]{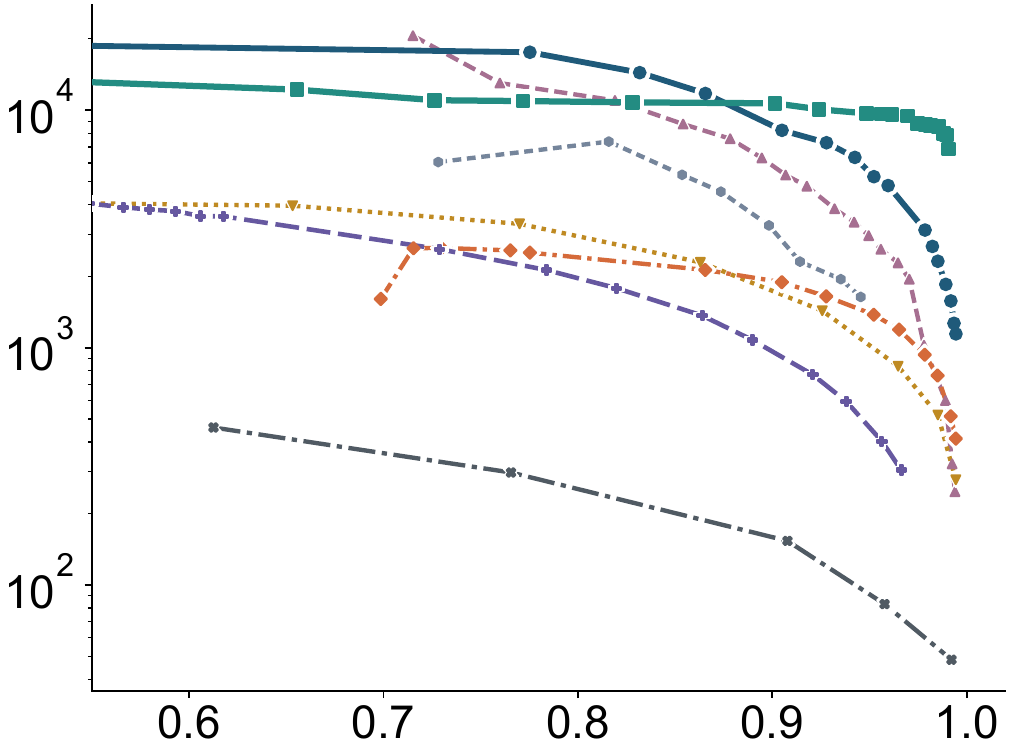}} &
    \raisebox{-0.5\height}{\includegraphics[width=0.228\textwidth]{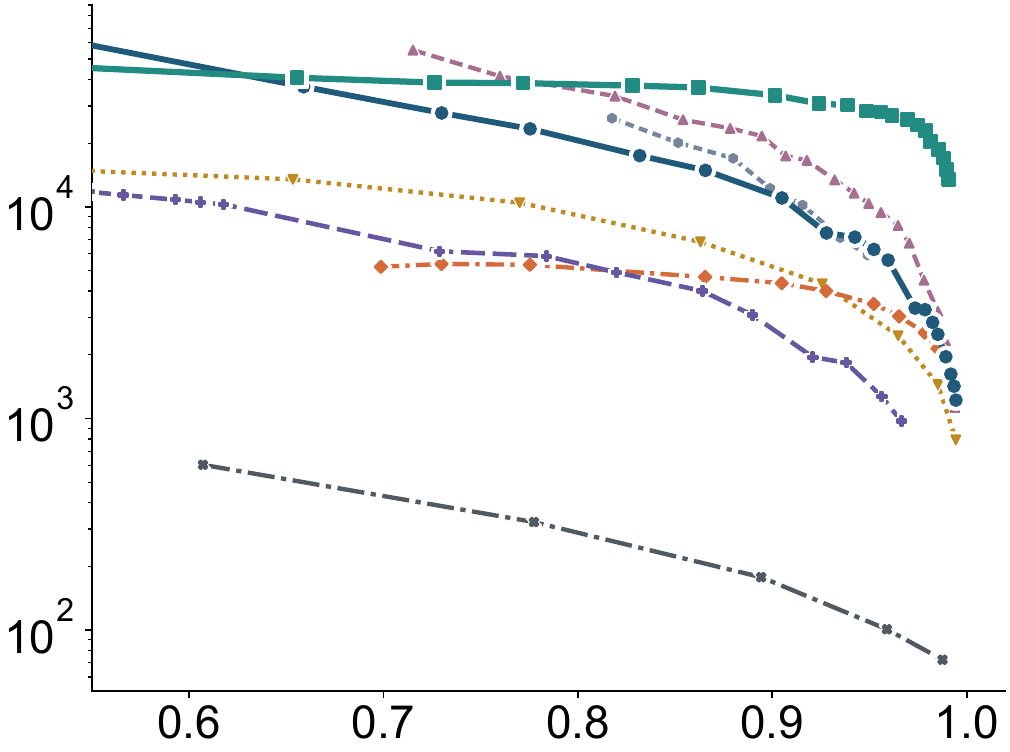}} &
    \raisebox{-0.5\height}{\includegraphics[width=0.228\textwidth]{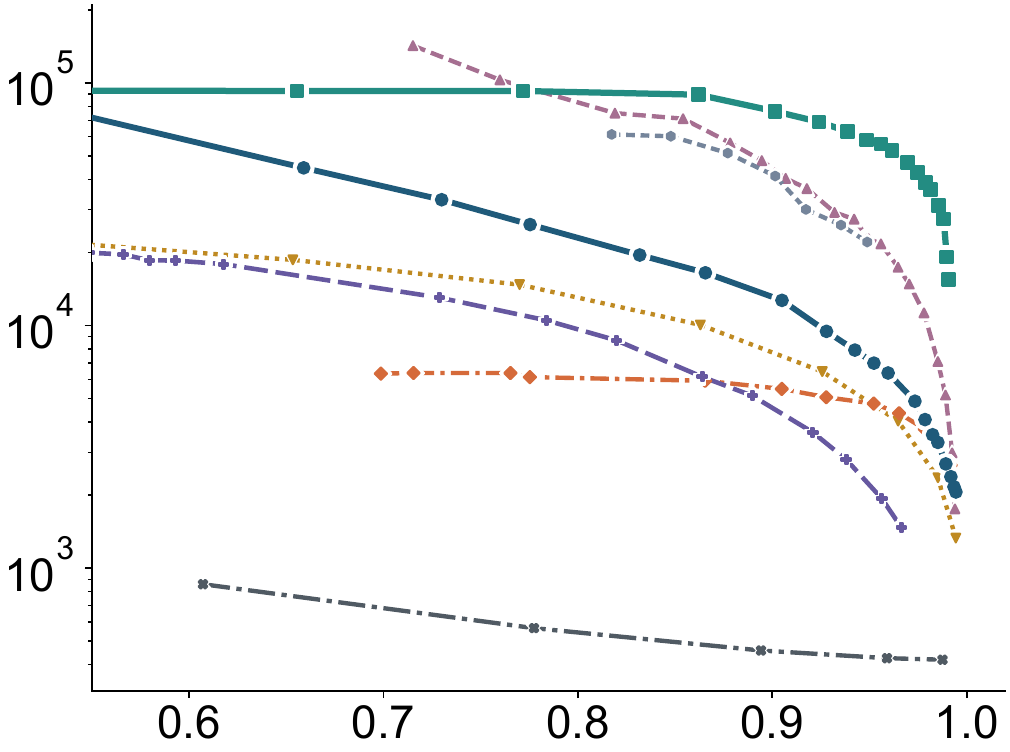}} &
    \raisebox{-0.5\height}{\includegraphics[width=0.228\textwidth]{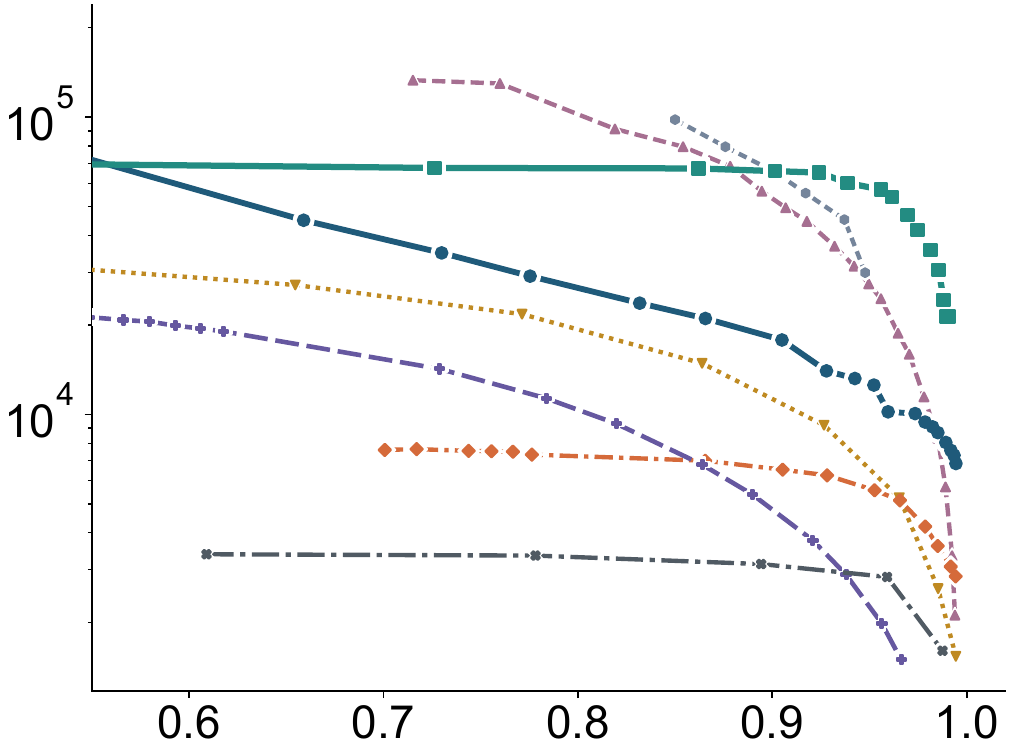}} \\[0.5em]
    \raisebox{-0.4\height}{\rotatebox{90}{\textbf{SPACEV-100M}}} &
    \raisebox{-0.5\height}{\includegraphics[width=0.228\textwidth]{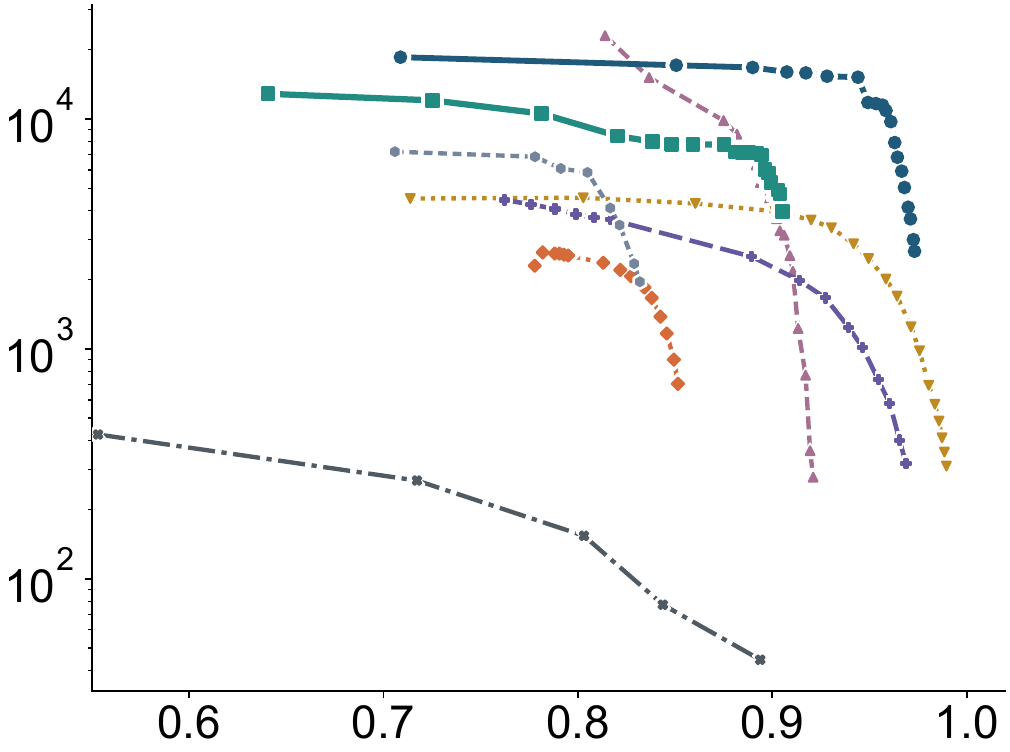}} &
    \raisebox{-0.5\height}{\includegraphics[width=0.228\textwidth]{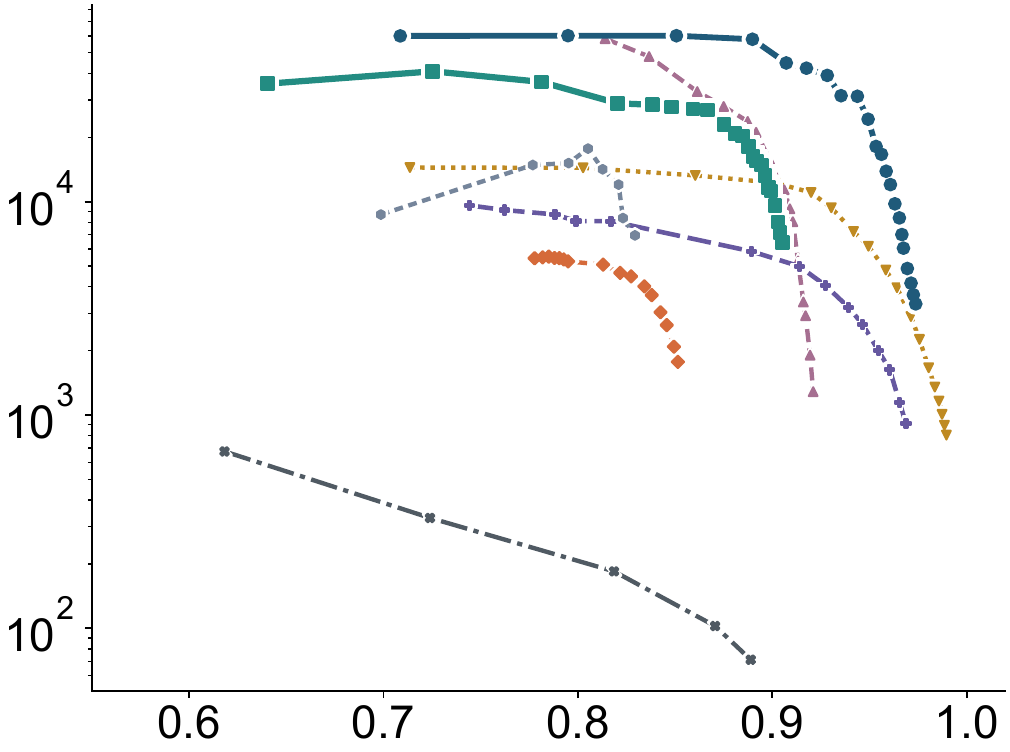}} &
    \raisebox{-0.5\height}{\includegraphics[width=0.228\textwidth]{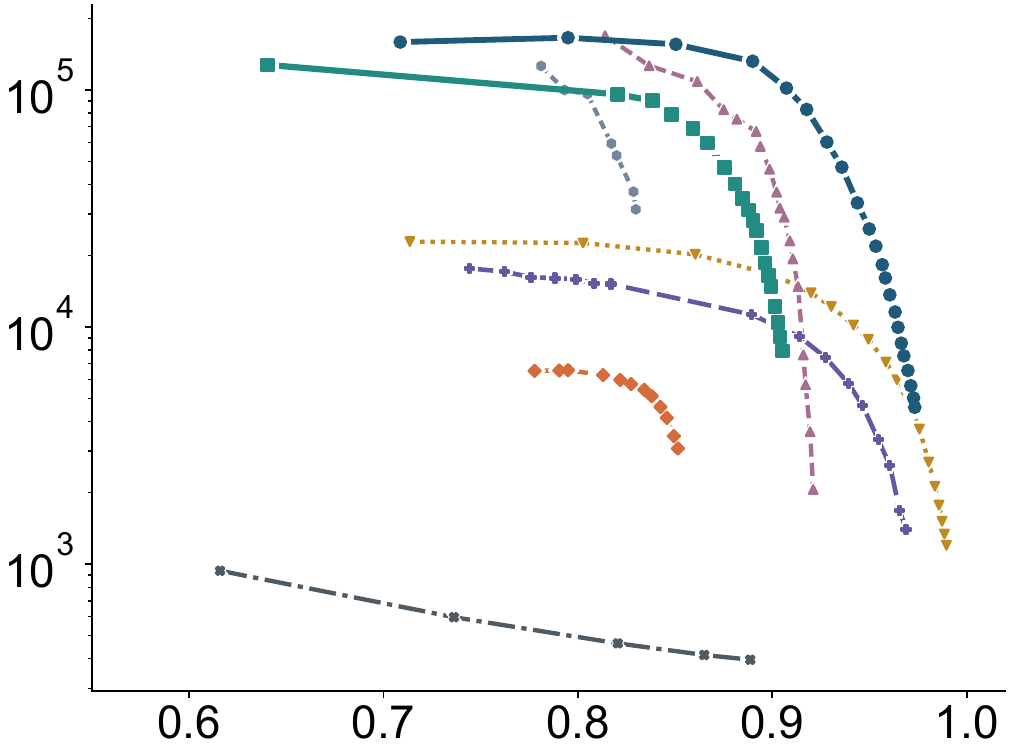}} &
    \raisebox{-0.5\height}{\includegraphics[width=0.228\textwidth]{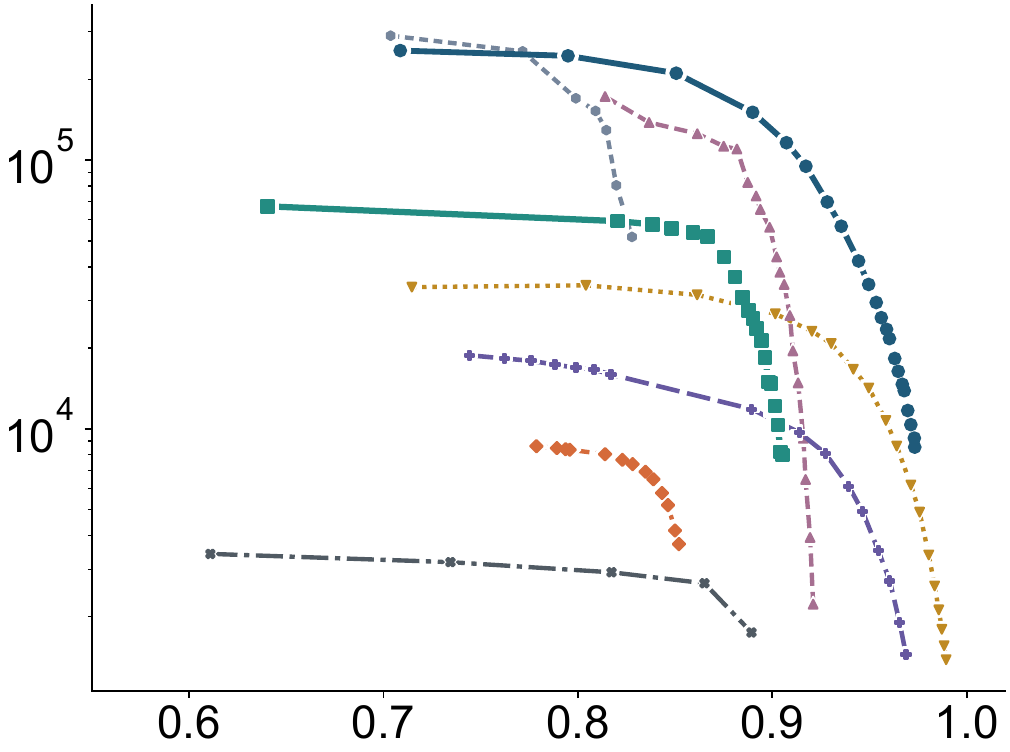}}
\end{tabular}%
}
\captionof{figure}{End-to-end Recall--QPS trade-offs on the 100M-scale datasets.}
\label{fig:graph_compare_100m}
\vspace{0.3em}

\begin{minipage}[t]{0.42\textwidth}
    \centering
    \includegraphics[width=\linewidth]{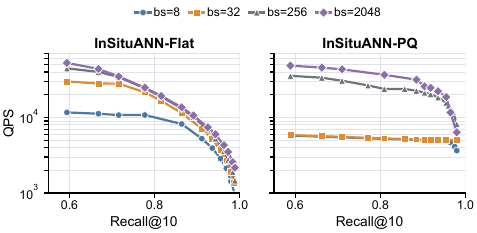}
    \captionof{figure}{Recall--QPS trade-off on the H100 node (SIFT-1B).}
    \label{fig:H100-main}
\end{minipage}
\hfill
\begin{minipage}[t]{0.42\textwidth}
    \centering
    \includegraphics[width=\linewidth]{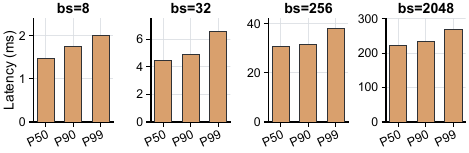}
    \captionof{figure}{\ourname{}-Flat batch-latency percentiles
    (Recall@10$\approx$0.9) on the H100 node.}
    \label{fig:h100-exact-latency-percentiles}
\end{minipage}
\vspace{0.3em}

    \includegraphics[width=0.92\textwidth]{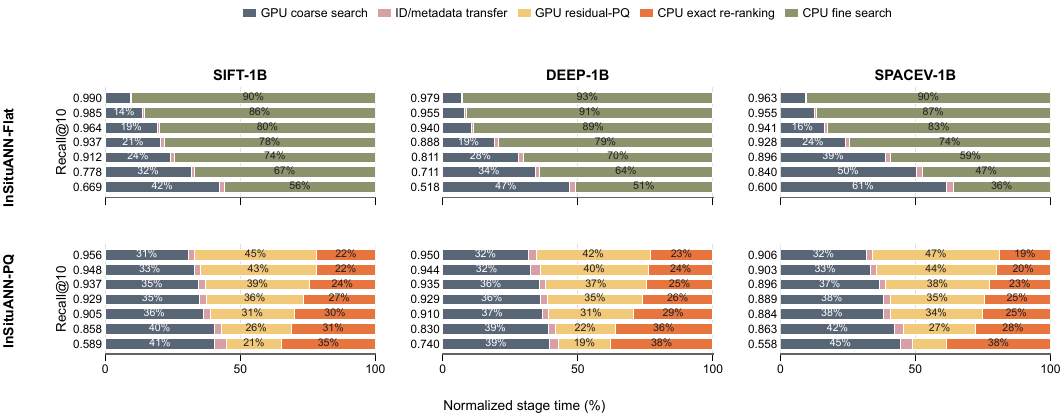}
    \caption{Normalized stage-time breakdown of \ourname{}-Flat and -PQ at
    batch size 8.}
    \label{fig:detailed-breakdown-bs8}
\end{figure*}

\end{document}